\documentclass[]{aa} % for a referee version
\usepackage{graphicx}
\usepackage{color}

\usepackage{longtable}
\begin{document}
%\title{XMM observation of Intracluster medium}
\title{Radial profiles of Fe abundance  in the intracluster medium of nearby clusters observed with  XMM-Newton}

\offprints{K. Matsushita}
\date{}
\author{Kyoko Matsushita}
\institute{Department of Physics, Tokyo University of Science, 1-3 Kagurazaka,
Shinjyuku-ku, Tokyo, 162-8601, Japan\\ \email{matusita@rs.kagu.tus.ac.jp}}
%\email{matusita@rs.kagu.tus.ac.jp}
\authorrunning{K. Matsushita}
\titlerunning{Fe abundance in the Intracluster Medium}
%Aims, Methods, and Results
\abstract
{}
{The abundances of Fe 
in the intracluster medium  of nearby ($z<0.08$) clusters were 
measured  up to 0.3$\sim$ 0.5r$_{180}$.
}{We analyzed 28 clusters of galaxies  observed with XMM-Newton.
We  derived Fe abundances from the flux ratios of Fe lines to the
continuum within an energy range of 3.5--6 keV to minimize and
evaluate systematic uncertainties.
}{  The radial profiles of the Fe abundances of
 relaxed clusters with  a cD galaxy at their X-ray peak  
     have similar slopes.  These clusters 
show similar enhancements in the Fe abundance within 0.1$r_{180}$,
and at 0.1--0.3$r_{180}$,
they have flatter 
Fe abundance profiles at 0.4$\sim$0.5 solar, with a small scatter.
Most  other clusters, including merging clusters,  
 also have similar Fe abundance profiles beyond 0.1$r_{180}$.
{  These clusters may have universal metal enrichment histories},
\rm
and a  significant amount of Fe  was
 synthesized at a very early stage  in  cluster formation. 
Mergers of clusters can destroy the central Fe peak.
}
{}
\maketitle

\section{Introduction}
Clusters of galaxies are the largest gravitationally bound objects in the universe.
The intracluster medium (ICM) contains a large amount of metals, which
are synthesized mainly by supernovae (SNe) in early-type galaxies
(e.g., Arnaud et al. 1992; Renzini et al. 1993).
Thus,  metal abundances in the ICM  provide important clues for
understanding the metal-enrichment history and evolution of galaxies in clusters.
Because both SN II and SN Ia synthesize Fe, 
the distribution of Fe in the ICM contains information about the star-formation histories of
massive stars and the history of chemical-enrichment  attributed to SN Ia. 

The ASCA satellite (Tanaka et al.\ 1994) first enabled us to measure the
distribution of Fe in the ICM\@ (e.g., Fukazawa et al.\ 2000; Finoguenov
et al.\ 2000; 2001).  
The Fe abundances of these clusters are 0.2--0.3 solar, adopting
the solar abundance from the ``photospheric'' values given by
Anders and Grevesse (1989). The
dependence of the Fe abundance
 on the temperature of the ICM is weak
 (Fukazawa et al. 1998).
ASCA revealed 
large-scale abundance gradients from AWM 7 and the Perseus cluster
(Ezawa et al. 1997; 2001).
De Grandi et al. (2004) derived the Fe abundances of nearby  hot clusters
observed with Beppo-SAX and  found that the abundance profiles are strongly
peaked for cool core clusters, whereas they remain constant for other systems.
 XMM-Newton and Chandra observations show a
spatial distribution of the Fe abundance of  up to $0.3\sim 0.4 r_{180}$
(e.g., Tamura et al. 2004; Vikhlinin et al. 2005; Baldi et al. 2007;
Maughan et al. 2008; Leccardi \& Molendi 2008).
 At the center of most relaxed clusters, the Fe abundance decreases steeply outward.
Outside the central region, the Fe abundance  %tight concentration and 
decreases more gradually toward the outer regions.

The Suzaku satellite first measured the Fe abundance of the ICM beyond $0.5r_{180}$.
(Fujita et al. 2008; Tawa 2008).  Within
0.3--0.5$r_{180}$, Suzaku can derive the Fe abundances more accurately
than XMM\@.  The Fe abundance  gradually decreases from the center to
$\sim $0.7$r_{180}$.

In this paper,  we describe our study of  the Fe abundance 
in the  ICM up to $0.3\sim 0.5$$r_{180}$
of   28 brightest clusters of galaxies observed with XMM-Newton.
Some clusters have cool cores at their center
(e.g. Makishima et al. 2001; B\"ohringer et al. 2002).
We must be careful when  deriving elemental abundances with a multi-temperature
plasma,  because emission lines and a continuum spectrum
 are different functions of temperature. 
Therefore, we
 directly derived the strength of the K$\alpha$ lines of Fe,
considering the temperature dependence of the flux ratios of 
the Fe lines and the  continuum,    and derived Fe abundances.
We adopted  an energy range in which the ICM dominates the background.

In Section 2 we summarize the observations and data preparation.
Section 3 describes our analysis of the data, and in Section 4 the
Fe abundances are determined.
We discuss our results in Secion 5.

We adopt  the new solar abundances  given by Lodders (2003),
according to which 
the solar  Fe abundance with regard to H is 2.95$\times 10^{-5}$by
number.
This  value differs  from the photospheric value (4.68$\times 10^{-5}$)
 given by  Anders and Grevesse (1989).
\color{black}
Considering the difference  in He abundance between the two solar abundance tables,
the Fe abundance yielded by 
the former is  1.5 times higher than that of the latter.
\color{black}
We use H$_0$ = 70 km/s/Mpc.
Unless otherwise specified, errors are quoted with 68\% confidence.

\section{Target selection and observations}

In the XMM-Newton archival data, we selected the 28 brightest clusters of galaxies
with z$<$0.08  and total  exposures above
 7 ks after filtering out background flares.
The samples are listed in Table 1.
The  ICM temperatures in the  
``the X-ray Brightest Abell Clusters (XBACs)''
by Ebeling et al. (1996)  range from 2.4 keV to 8.7 keV.
Seventeen clusters show relatively relaxed morphologies with a cD galaxy 
at the X-ray peak. Hereafter, these are called cD clusters.
Eleven clusters are merging clusters or relaxed clusters with two 
giant galaxies in the  central region such as the Coma cluster.
Hereafter, these clusters are called non-cD clusters.

The virial radius of each cluster, $r_{180}=1.95
h_{100}^{-1}\sqrt{k<T>/10 \rm{keV}}$,
was calculated following Markevitch et al. (1998), by adopting an  average
temperature  $<T>$  of  that at
  0.06-0.3$r_{180}$ from the X-ray peak of each cluster.

We used the PN, MOS1, and MOS2 detectors  to derive the Fe abundance of the ICM.
We selected events with patterns smaller than \color{black} 5 and 13 \color{black} for the PN 
and MOS, respectively.

\begin{figure*}[]
\resizebox{17cm}{!}{\includegraphics{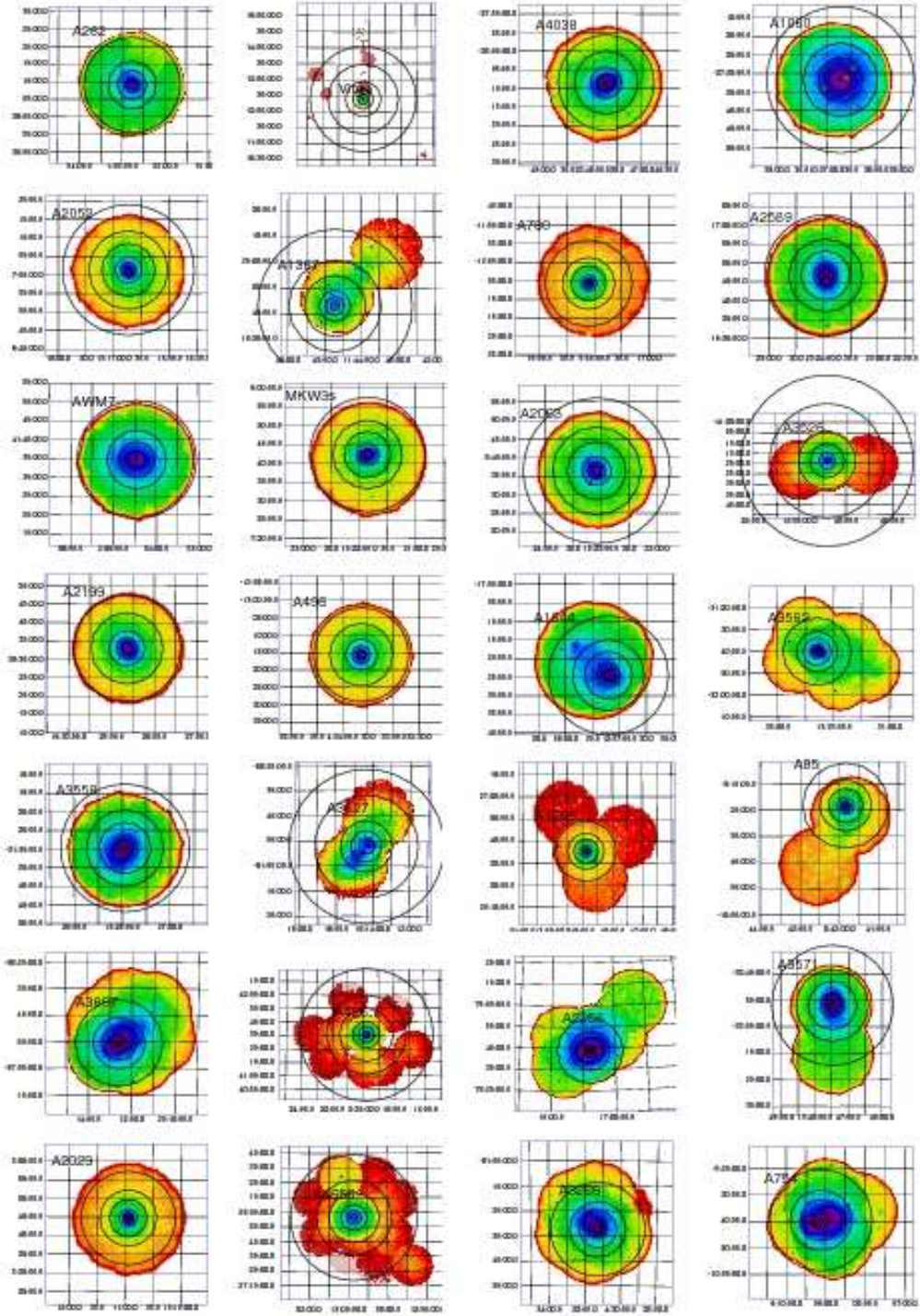}}

\caption{Combined MOS and PN images of the target clusters. 
Exposures are not
 corrected and therefore the  exposures of each field can be visualized.
 Circles have radii of 
0.03, 0.06, 0.1, 0.2, 0.3, and 0.5$r_{180}$. %and small
}
\label{images}
\end{figure*}

Spectra were  accumulated in concentric annular regions of
 0--0.03$r_{180}$, 0.03--0.06$r_{180}$, 0.06--0.1$r_{180}$, 0.1--0.2$r_{180}$, 0.2--0.3$r_{180}$, and
0.3--0.5$r_{180}$
centered on the X-ray peak of each  cluster.
Here, the X-ray peaks were derived using the  ewavelet tool of SAS-v8.0.0,
and luminous point sources were excluded.
The calculated X-ray peak of each cluster
and the accumulation regions for the spectra are summarized
as X-ray images in  Figure \ref{images}.
The  spectra from  MOS1 and MOS2 were added.
The background spectrum was calculated 
by  integrating  blank-sky data in the same detector regions.
From deep-sky observations with the XMM, we selected the data
with the  background most similar to that of each cluster,
 and the faintest Galactic emission,
after screening out  background flare events in the data and the background,
following Katayama et al. (2004).
Then, the background was scaled using the count rate between 10 and 12 keV.
\color{black} Most scaling factors are  smaller than  10\%,  because
we selected a blank-sky observation to match each cluster observation.
\color{black}

The response matrix file  and the auxiliary response file (ARF) corresponding
to each spectrum were calculated using  SAS-v8.0.0.
We used the XSPEC\_v11.3.2ag package for our spectral analysis.

\begin{table*}[thb]
\caption{Cluster samples in the XMM-Newton archival data}
 \begin{tabular}[t]{llrrrllll}
\hline \hline
%\scriptsize
cluster & \multicolumn{2}{c}{z ~ $<T>^a$}& $N_{\rm{H}}^b$ &type$^c$ & obsid (exposures of PN, MOS1 and MOS2)$^d$\\
 & & (keV) &\multicolumn{2}{l}{(10$^{20}\rm{cm^{-2}}$)}& (ks, ks, ks)\\
\hline
A262 & 0.016 & 2.5 & 5.4 & cD &  0109980101 (19.0  23.6  23.9)   0504780201 (24.9  31.6  33.1)\\
Virgo & 0.004 & 2.8 & 2.6 & cD & 0114120101 (30.0  37.7  38.5) 0200920101 (69.6  76.9  78.5)
\\ &&&&& 0106060101 (5.0  9.2  9.2) 0106060201 (1.6  5.5  5.0)
 0106060301 (5.4  7.1  7.9)\\ &&&&& 0141570101 (17.4  23.2  24.7) 0112550701 (1.0  0.2) 0112552101 (7.6  13.4  13.5) \\ &&&&& 0106060401 (7.3  10.7  11.1) 0145800101 (5.9  37.8  0.2  0.2) 0106060501 (10.9  15.3  15.4)\\ &&&&& 0200650101 (42.7  53.2  53.5) 0112840101 (14.6  18.1  17.6) 0106060601 (8.6  10.5  10.7) \\ &&&&&0112550801 (2.9  13.2  12.1) 0106060701 (1.2  12.4  5.9) 0108860101 (16.7  20.6  19.2) \\ &&&&& 0106061401 (4.9  8.0  8.3) \\
A4038 & 0.029 & 3.0 & 1.5 & cD &  0204460101 (27.8  29.3  28.9)\\
A1060 & 0.012 & 3.0 & 4.9 & non-cD &  0206230101 (28.5  37.9  38.9)\\
A2052 & 0.035 & 3.1 & 3.0 & cD &  0109920101 (26.4  28.9  29.0) \\
A1367 & 0.021 & 3.3 & 2.4 & non-cD &  0061740101 (23.7  31.8  31.4) 0005210101 (20.8  27.8  28.8)\\
A780 & 0.057 & 3.5 & 4.7 & cD &  0109980301 (14.0  19.1  22.2)\\
A2589 & 0.042 & 3.6 & 3.9 & cD &  0204180101 (21.8  26.3  27.5)\\
AWM7 & 0.017 & 3.6 & 9.8 & cD &  0135950301 (27.5  30.7  30.9)\\
MKW3s & 0.045 & 3.7 & 3.0 & cD &  0109930101 (0.0  37.7  38.0)\\
A2063 & 0.036 & 3.9 & 3.0 & cD &  0200120401 (5.0  8.2  5.0)\\
A3526 & 0.010 & 4.0 & 8.2 & cD &  0046340101 (35.2  44.9  43.5) 0406200101 (90.5  111.7  111.7) 0504360101 (26.4  28.2  30.9) \\  &&&&&  0504360201 (32.9  34.0  34.6)\\
A2199 & 0.030 & 4.2 & 0.9 & cD &  0008030601 (0.5  6.9  4.7) 0008030301 (2.9  4.7  4.7) 0008030201 (13.3  14.5  14.6)\\
A496 & 0.033 & 4.4 & 4.6 & cD &  0135120201 (11.8  16.2  15.9)  0506260301 (44.5  57.0  59.5) 0506260401 (44.8  55.8  58.6) \\
A1644 & 0.048 & 4.6 & 4.8 & non-cD &  0010420201 (12.7  14.1  14.1) \\
A3562 & 0.050 & 4.8 & 4.0 & non-cD &  0105261301 (37.7  39.0  39.5) 0105261501 (4.8  22.4  22.4) 0105261801 (4.7  10.9  10.4)\\ &&&&&  0105261701 (0.0  20.6  20.4) 0105261601 (16.3  18.3  19.5) 0105261401 (9.0  12.8  12.4)\\
A3558 & 0.048 & 5.4 & 4.2 & non-cD &   0107260101 (38.2  43.1  42.5)\\
A3627 & 0.016 & 5.5 & 20.8 & non-cD &  0204250101 (2.4  6.2  5.8) 0208010101 (9.1  12.6  13.0) 0208010201 (10.9  13.6  14.3)\\
A1795 & 0.062 & 5.8 & 1.2 & cD &  0097820101 (25.1  38.3  37.9) 0109070201 (53.0   54.5    54.6) 0205190101 (26.9  29.5  28.8)\\ &&&&&  0205190201 (20.8  22.6  23.2)\\
A85 & 0.052 & 5.8 & 3.1 & cD &  0065140101 (9.8  12.4  12.2) 0065140201 (9.2  12.1  12.4)\\
A3667 & 0.053 & 6.0 & 4.6 & non-cD & 0105260101 (0.8  7.3  6.8) 0206850101 (52.3  59.2  60.0) 0105260601 (17.7  23.1  23.3) \\ &&&&& 0105260301 (12.8  17.0  16.2) 0105260401 (11.5  16.3  16.5) 0105260501 (11.2  12.6  12.6) \\ &&&&& 0105260201 (15.1  19.0  18.4)\\
A426 & 0.018 & 6.1 & 14.6 & cD & 0085110101 (47.5  48.8  50.7) 0204720101 (11.9  15.1  15.2) 0085590201 (38.4  41.7  40.0) \\(Perseus) &&&&& 0305720101 (10.0  13.0  12.6) 0305720301 (15.9  19.9  21.2) 0151560101 (18.0  24.9  25.1) \\ &&&&& 0204720201 (20.7  24.5  24.5) 0405410201 (9.4  17.3  25.5) 0405410101 (12.2  18.2  16.1)\\
A2256 & 0.058 & 6.3 & 4.1 & non-cD &  0141380101 (6.4  9.2 8.7) 0141380201 (10.6  13.2  12.4 )  \\ &&&&&  0112951501 (5.9  10.6  10.9) 0112951601 (7.2  12.7  12.7)  0112950601 (8.4  12.3  11.8) \\ &&&&&   0112500101 (21.8  25.0  25.2)\\
A3571 & 0.040 & 6.5 & 4.4 & cD &  0086950201 (14.3  25.4  25.2) \\
A2029 & 0.077 & 7.4 & 3.2 & cD &  0111270201 (9.3  12.1  12.4)\\
%\end{tabular}
%\end{table*}
%\setcounter{table}{0}
%\begin{table*}[thb]
%\caption{(continue)}
%%\vspace*{-0.5cm}
% \begin{tabular}[t]{llrrrllll}
A1656 & 0.023 & 7.8 & 0.9 & non-cD &0300530201 (0.4  3.6   3.4 ) 0153750101 (19.2  21.3  21.8) 0124711401 (14.7  17.4  16.5) \\ (Coma)&&&&& 0300530101 (18.6  20.9  21.0) 0300530301 (26.5  30.5  30.2) 0300530701 (16.4  24.4  24.6) \\ &&&&& 0300530401 (13.3  18.8  19.6) 0300530601 (16.7  22.6  22.1) 0300530501 (19.2  24.7  24.6)\\ &&&&& 0124710901 (16.7  21.6  22.1) 0124712001 (9.2  15.4  14.6) 0124710501 (10.7  24.4  24.8) \\ &&&&& 0124710601 (4.7  9.5  9.0) 0204040101 (69.7  73.0  74.0) 0204040301 (43.7  52.9  53.2)\\ &&&&& 0204040201 (54.8  61.8  64.8)  0124710801 (18.4  24.5  24.5) 0124710401 (8.2  5.9  5.6) \\ &&&&& 0124712501 (23.8  27.7  27.7) 0124712401 (11.0  17.4  18.1) 0124711101 (13.2  19.3  21.1)\\ &&&&& 0124712201 (18.2  26.7  26.2) 0124712101 (20.3  26.0  26.1)  0124710301 (10.6  13.4  14.0)\\ &&&&& 0124710101 (23.7  30.6  30.8)\\
A3266 & 0.055 & 8.4 & 1.5 & non-cD &0105262201 (3.4  0.0  3.6) 0105260901 (18.6  24.4  24.3) 0105260801 (16.2  19.6  20.4)\\ &&&&&  0105262501 (3.5  7.6  7.7)  0105262001 (2.9  7.6  7.6) 0105260701 (16.1  20.6  20.5) \\ &&&&& 0105261101 (8.0  12.7  11.1) 0105262101 (4.4  5.6  7.2)\\
A754 & 0.054 & 8.6 & 4.7 & non-cD &  0112950301 (8.0  13.3  13.7) 0112950401 (7.6  14.3  14.3) 0136740201 (2.3  6.0  6.7) \\ &&&&& 0136740101 (12.5  14.2  14.2)\\
\hline

 \end{tabular}\\
\\
$^a$:Adopted average  temperature of ICM used to calculate $r_{180}$.\\
$^b$:The Galactic value of the hydrogen column density \color{black}(Dickey \& Lockman 1990)\\\color{black}
$^c$: ``cD''  corresponds to cD clusters, which are
 relaxed clusters with a cD galaxy at their
 center. Other clusters are  classified as ``non-cD.''\\
$^d$:obsid (a unique XMM-Newton observation identifier) 
and  exposure times of PN, MOS1, and MOS2,
respectively, after screening out the background flares.
\end{table*}

%\clearpage
\section{Analysis }

\subsection{Fe line strength}\label{feline}

In this study, the Fe abundances of the ICM are
 derived from the ratio of the flux  in units of
photons $\cdot\rm{cm}^{-2}\rm{s}^{-1}$  of the K$\alpha$ lines of 
 He-like Fe (hereafter $F_{\rm FeHeK}$) or H-like Fe (hereafter $F_{\rm
 FeHK}$) to that of  the continuum  at 3.5--6.0 keV (hereafter
 $F_{3.5-6}$), 
since the  systematic uncertainty in the continuum flux in this energy 
band owing to  the background  is  smaller than that around the Fe-K lines.
Another advantage is that 
 the  dependence of the ratios  on the  plasma temperature 
is relatively weak within a certain temperature range.
The temperature dependence of these ratios
  can  constrain  on the  Fe abundances of multi-temperature plasmas.

\begin{figure}[t]
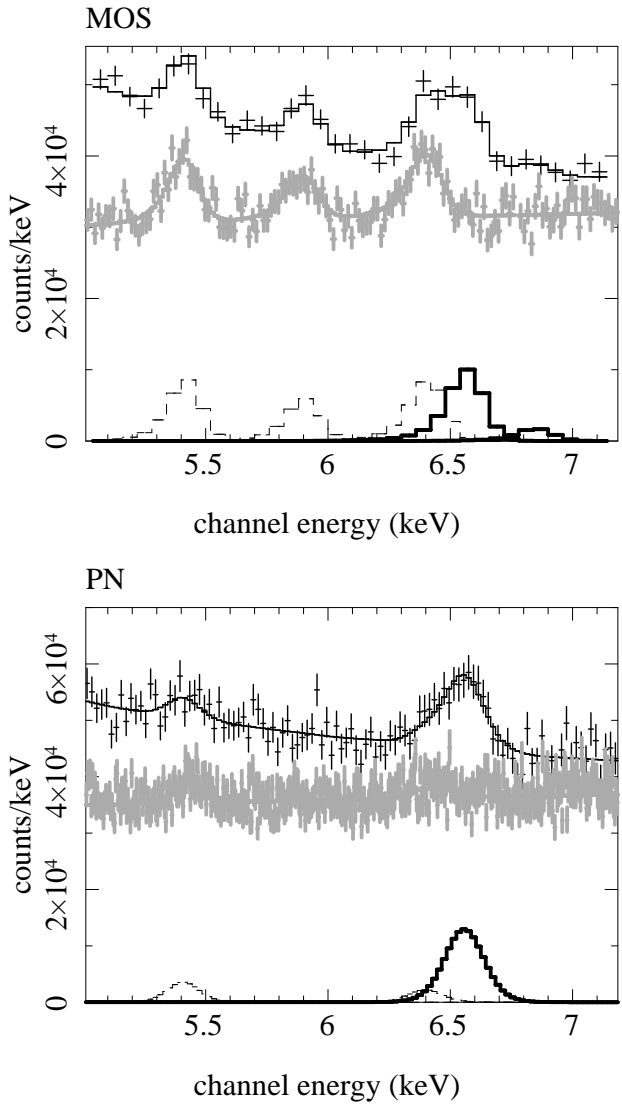

%\resizebox{8.1cm}{!}{\includegraphics{plot8/perse_off_mos.ps}}
%\resizebox{8.1cm}{!}{\includegraphics{plot8/perse_off_pn.ps}}
\resizebox{8.1cm}{!}{\includegraphics{13432fig2a.ps}}
\resizebox{8.1cm}{!}{\includegraphics{13432fig2b.ps}}
%\resizebox{8.1cm}{!}{\includegraphics{13432fig2b.ps}}
\caption{
Representative spectra of Abell 426 (the Perseus cluster)  (black)
and the adopted background (gray) at
%0.06--0.1$r_{180}$ (MOS:red,  PN:magenta), and
0.3--0.5$r_{180}$ from  MOS and PN fitted with
Gaussians and continuum. % \color{black} ????$B"r!-(BMOS,pn labeling \color{black}
Best-fit Gaussians corresponding to the K$\alpha$-lines of He-like and H-like Fe
are plotted as bold  lines: those for background lines are plotted
as thin dashed lines.
}
\label{FespecA426}
\end{figure}

To obtain suitable  statistics,
we simultaneously fitted the raw annular spectra of a cluster   and a deep field 
 within an energy range of 5.0--7.2 keV 
with the sum of the ICM and background emission.
The ICM component consists of 
 thermal bremsstrahlung and two Gaussians for the K$\alpha$ lines of
He-like and H-like Fe.
We modeled the background emission with 
 a power-law model for the cosmic X-ray background (CXB), a  ``powerlaw/b'' model for %remaining
non X-ray background (NXB), and three Gaussians at 5.4 keV,
5.9 keV, and 6.4 keV, \color{black} which are the K$\alpha$ lines of 
neutral  Cr, Mn, and Fe, respectively. 
\color{black}
The "powerlaw/b"  model is not folded through the ARF, and
differs from a power-law.
\color{black}
We obtained acceptable fits with a reduced $\chi^2\sim 1$.
Figure \ref{FespecA426} shows representative spectra of Abell 426
(A426, the Perseus cluster)
fitted in this way.
These spectra are fitted well, and
 the  K$\alpha$ line of He-like Fe is clearly detected even at
 0.3--0.5$r_{180}$.
 The  background line at 6.4 keV may cause a systematic uncertainty 
in the strength of the He-like Fe line in the  MOS spectra at low-brightness regions.
In contrast, 
the background lines in the PN spectra in this energy range are much weaker
than those in the MOS.

Despite the 6.4 keV background line,
 the PN and MOS  gave mostly consistent 
  ratios of  $F_{\rm FeHeK}$/$F_{3.5-6}$ and  $F_{\rm
FeHK}$/$F_{3.5-6}$.
Hereafter
 we used the weighted average of these ratios from the PN and MOS
to derive Fe abundances with relative errors smaller than 30\%.

\subsection{Spectral fit with single-temperature model}

To convert the observed line flux to the Fe abundance, 
we derived the ICM temperatures.
The first step in this analysis was to fit the 
background-subtracted
annular (projected)
spectra from the PN and MOS with a single-temperature vAPEC model (\cite{apec}) with
 photoelectric absorption fixed at the Galactic value (hereafter the 1T model).
The energy ranges of 1.4--1.6 keV for the MOS and 7.2--9.0 keV for the PN
were discarded owing  to   strong instrumental lines.
To account for the remaining background,
we added a ``powerlaw/b'' model 
(Zhang et al. 2009) and fitted the spectra 
within the energy range of \color{black}1.2\color{black}--10.0 keV.
Here, we 
 fixed the photon index of the ``powerlaw/b'' component 
at 0.15 and 0.35 for the MOS and PN, respectively (Zhang et al. 2009).
The abundances of C, N, and Al were fixed at 1 solar and \color{black} those of
O and Ne were assumed to have the same values as Mg.
The abundances  of Mg, Si, S, Ar, Ca, Fe, and Ni were allowed to vary.
\color{black}
We did not use a more detailed modeling of the background (Snowden et
al. 2008; Leccardi \& Molendi 2008), since we mainly used an energy range 
over which the ICM emission is dominant.
The ICM temperatures derived from the PN and MOS
are consistent with each other within the order of several percent.
We adopted the weighted average of the temperatures 
derived from the PN and MOS, as 
summarized in Table \ref{fetable}.

\color{black}
To derive the Fe abundance in each annular region,
 we included a 10\% systematic uncertainty in temperatures,
considering the  uncertainties described below.
When the spectra are fitted within the energy range of 0.8--10.0 keV,
which 
 includes Fe-L emission, 
the derived temperatures decreased systematically  by $\sim$10\%.
We also fitted the spectra without the ``powerlaw/b'' model.
Then, outside 0.2$r_{180}$,
the derived ICM temperature  increased by  10$\sim$20\%  and the
discrepancy between the PN and MOS became increased.
Similarly, we
 also analyzed the same XMM data in the same way with SAS v7.0.0.
In this case,
%However,
 when the derived temperatures are   higher than  several  keV, 
 the MOS systematically gives
$\sim$10\% higher than those given by the PN.
Systematic differences of  $\sim$10\% in cluster temperature 
among Chandra,  the PN, and the MOS were reported
in relatively high-temperature clusters (Snowden et al. 2008).
Reese et al. (2010) found that the ICM temperature derived from Chandra
data changes systematically by $\sim$10 \% between calibrations.
\color{black}

\color{black}

\subsection{Spectral fit with multi-temperature model}

To study the effect of the temperature structure of the ICM,
we also fitted the background-subtracted spectra of  the  MOS 
 at \color{black}0.5-10.0 keV\color{black},
using   a multi-temperature vAPEC model with photoelectric
absorption (multi-T model), the Galactic emission, and the ``powerlaw/b'' model with 
a fixed photon index, as  described in Section  3.2. 
\color{black}
The Galactic emission, which includes the local hot bubble,
the Milky Way halo, and solar wind charge exchange,
 is empirically fitted with a two-temperature
APEC model with redshift $=0$ (e.g. Lumb et al. 2002; Yoshino et al. 2009 ).
Therefore, we used the two-temperature APEC model for the Galactic emission.
The temperature of the one component was fixed at 0.1 keV,
and that of the other  was restricted within 0.2--0.4 keV.
\color{black}
Here, we used only the MOS detector,
because  it has a better
energy resolution and is more sensitive to temperature structure
than   the PN detector. 
The multi-T model is a sum of 13 temperature components,
because the spectrum of  an isothermal plasma can be reproduced well  by the sum of
the spectra of two neighboring temperature plasmas with similar elemental
abundances.
The ratio of the temperatures of the two neighboring components is fixed at 0.8.
The hottest temperature of the multi-T model is fixed at  
the highest temperature derived using  the 1T model 
derived from the MOS within 0.2$r_{180}$.
The elemental abundances of each element for the multi-T 
components are assumed to have the same values.
We   added a 3\% systematic error to the count rate of each channel in
the spectra. 
Then, the reduced $\chi^2$ reaches the reasonable values of $\sim$1.

\section{Results}

\subsection{Temperature profiles of the ICM}

\begin{figure}[tbh]
\resizebox{9cm}{!}{\includegraphics{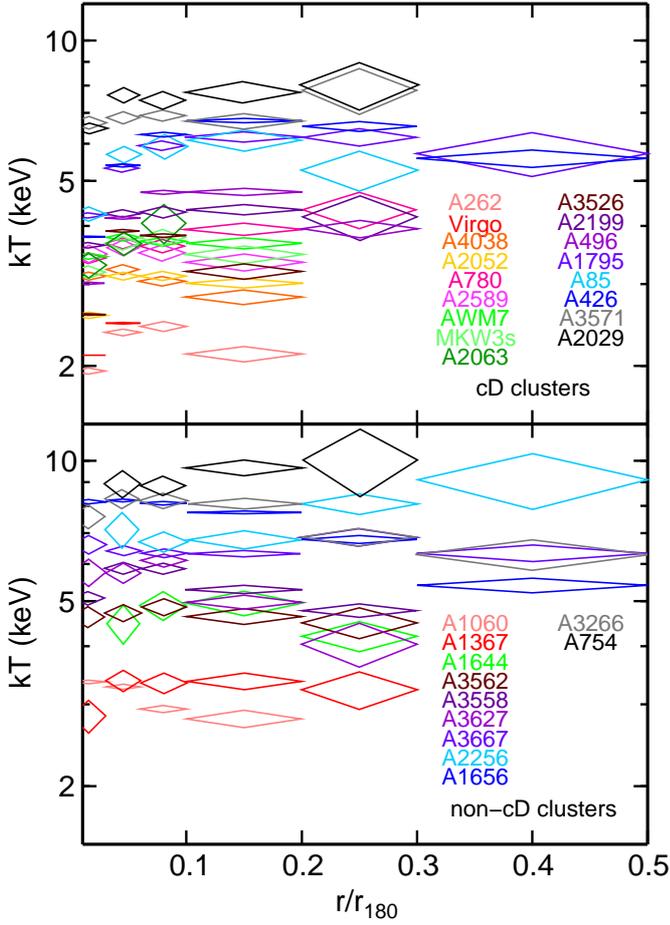}}
\caption{\color{black}
Radial profiles of the ICM temperature of  cD clusters
(upper panel) and non-cD clusters (lower panel).
}
\label{kTr}
\end{figure}

The radial temperature profiles of the ICM  are summarized in Figure \ref{kTr}.
From 0.06$r_{180}$ to 0.2$\sim$0.3 $r_{180}$, the temperature gradient of the ICM
is flat, whereas the temperatures of several cD clusters
are lower within 0.06$r_{180}$ 
 than those at 0.06--0.2$r_{180}$, owing  to the presence of cool
 cores.
In contrast,
 most  non-cD clusters  have  flatter radial temperature profiles
within 0.2$\sim 0.3 r_{180}$.
Beyond 0.2$\sim$0.3$r_{180}$, the ICM temperatures tend to decrease  outward.
These temperature profiles are consistent with previous 
Chandra and XMM-Newton measurements (e.g., Snowden et al. 2008; Vikhlinin
et al. 2005).

\subsection{Fe abundance of the ICM}

\begin{figure}[tbh]
\resizebox{8.1cm}{!}{\includegraphics{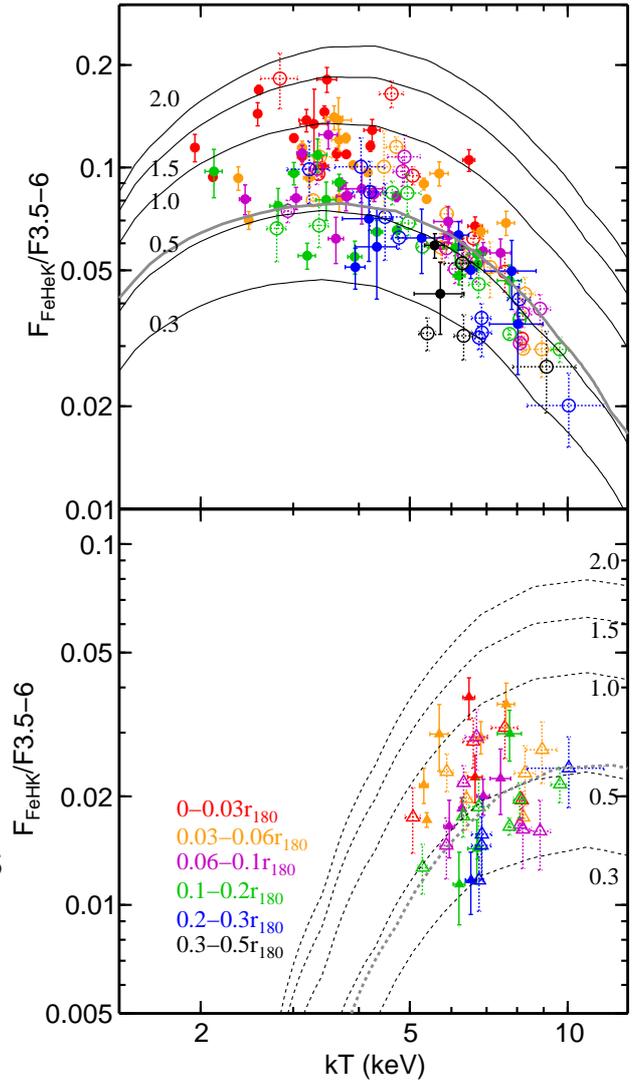}}
\caption{
 $F_{\rm FeHeK}/F_{3.5-6}$ (\color{black}upper panel; circles) and 
 $F_{\rm FeHK}/F_{3.5-6}$ (\color{black}lower panel; triangles)
 at 0--0.03$r_{180}$ (red),
0.03-0.06$r_{180}$ (orange), 0.06-0.1$r_{180}$ (magenta)
 0.1-0.2$r_{180}$ (green), 0.2--0.3$r_{180}$ (blue),  and 0.3--0.5$r_{180}$ (black),
 plotted against the ICM temperature. % derived from the 
Closed and open symbols indicate cD clusters and non-cD clusters,
 respectively.
Black solid and dotted lines correspond to the theoretical ratios from
the APEC model 
of  $F_{\rm FeHeK}/F_{3.5-6}$ and  $F_{\rm FeHK}/F_{3.5-6}$,
 respectively, assuming  \color{black}
 Fe abundance = 0.3, 0.5, 1.0, 1.5, and 2.0 solar.
Here, abundance ratios of elements are assumed to have the solar ratios
by Lodders (2003). 
\color{black}
Those of the MEKAL model assuming Fe abundance = 0.5 solar are plotted as gray lines.
}
\label{feratio}
\end{figure}

Figure \ref{feratio} shows the observed values of  $F_{\rm
 FeHeK}/F_{3.5-6}$ and $F_{\rm FeHK}/F_{3.5-6}$, plotted against
the ICM temperature derived from the 1T model. 
Within $0.03 r_{180}$,  $F_{\rm FeHeK}/F_{3.5-6}$ 
is scattered from 0.03 to 0.2, with  mild temperature dependence.
Beyond  $0.03 r_{180}$,  
both  $F_{\rm FeHeK}/F_{3.5-6}$ and $F_{\rm FeHK}/F_{3.5-6}$
become closer to the theoretical relationships predicted by the APEC model  assuming
that the  Fe abundance=0.5 solar.
Beyond $0.3r_{180}$, the data for only several clusters have adequate statistics:
the  $F_{\rm FeHeK}/F_{3.5-6}$ values of A426 (the Perseus cluster) and a non-cD
cluster are close to the relationship for 0.5 solar, whereas
those of two non-cD clusters are significantly lower.

The temperature dependence of the
theoretical ratio   $F_{\rm FeHeK}/F_{3.5-6}$ for 
a given Fe abundance is fairly weak,
 within 20\%  for 2--6 keV,
and  starts to decrease  above 6 keV (Figure \ref{feratio}).
In contrast, the ratio of $F_{\rm FeHK}/F_{3.5-6}$
 is nearly constant within 20\% for  6.7--17 keV.
This temperature 
independence  minimizes the systematic
 uncertainty in the Fe abundance because of  uncertainties in the
temperature structure.

To  derive  the Fe abundance, we  \color{black}converted \color{black} the observed
ratio of $F_{\rm FeHeK}/F_{3.5-6}$  using the theoretical 
ratios calculated by the APEC model and  the ICM temperature derived from the 1T model.
The results are summarized in Table \ref{fetable}.

\subsection{Radial profiles of  Fe abundances}

\begin{figure}[tbh]
\resizebox{8cm}{!}{\includegraphics{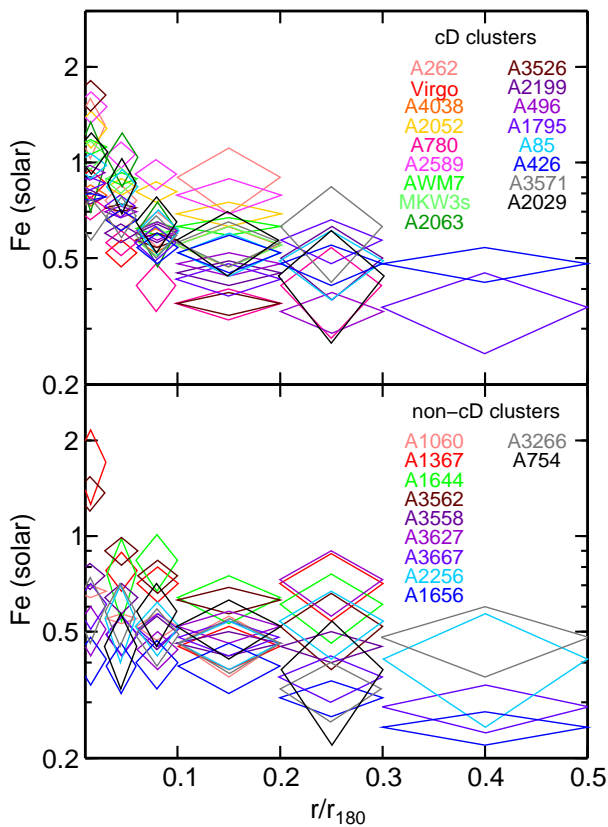}}
\caption{\color{black}
Radial profiles of the Fe abundances of  cD clusters
(upper panel) and non-cD clusters (lower panel).
}
\label{Fer}
\end{figure}
\begin{figure}[tbh]
\resizebox{8cm}{!}{\includegraphics{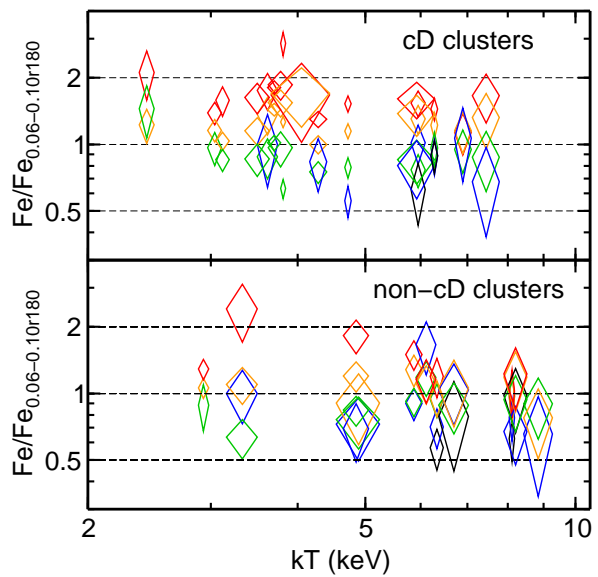}}
\caption{\color{black}
Ratios of Fe abundances derived from
  $F_{\rm FeHeK}/F_{3.5-6}$ 
at 0--0.03$r_{180}$ (red),
0.03-0.06$r_{180}$ (orange),
 0.1-0.2$r_{180}$ (green), 0.2--0.3$r_{180}$ (blue),  and 0.3--0.5$r_{180}$ (black) to 
 those at 0.06--0.1$r_{180}$, plotted against the ICM temperature
at 0.06--0.1$r_{180}$.
Upper and lower panels
correspond to the cD and non-cD clusters,
respectively. 
}
\label{feabundr}
\end{figure}

The radial  profiles of the  Fe abundances in the ICM are shown in Figure 
\ref{Fer}.
\color{black}
Within $0.03r_{180}$, the derived Fe abundances
\color{black} of the cD clusters are 0.7--1.6 solar. 
Non-cD clusters with  large deviations from spherical
symmetry,  A1367 and A3562, also have high Fe abundances of $\sim 1.5$ solar.
In contrast,
the other   non-cD clusters   have  lower { \color{black}  central values of Fe}.
Outside 0.1$r_{180}$, both the cD and non-cD clusters have similar radial profiles
of the Fe abundance.
At  0.1--0.2$r_{180}$, the derived Fe abundances are $\sim$0.5  solar.
For the region at 0.3--0.5$r_{180}$, 
although sufficient statistics are available for   only six  clusters,
  the derived Fe abundances are  around 0.3--0.5 solar.

\color{black}
Figure \ref{feabundr} shows the ratios of the Fe abundances of each cluster
at each annular region to that at 0.06--0.1$r_{180}$. 
Except for the  Centaurus cluster (A3526),
the Fe abundances of the  cD clusters have similar slopes:
the Fe abundances within 0.03$r_{180}$ are higher than 
those at 0.06-0.1$r_{180}$ by a factor of 1.5.
Beyond 0.06$r_{180}$,  the radial profiles become flatter.

\color{black}
The Fe abundances of
 three morphologically distorted clusters, A1367, A3558, and A3562,
within 0.03$r_{180}$ are higher than those  at 0.06-0.1$r_{180}$
by a factor of 1.5--2.
The other non-cD clusters have similar Fe abundances within 0.03$r_{180}$ and 
0.06-0.1$r_{180}$.

\subsection{Temperature dependence of the Fe abundance}

\color{black}

\begin{figure}[tbh]
\resizebox{8.1cm}{!}{\includegraphics{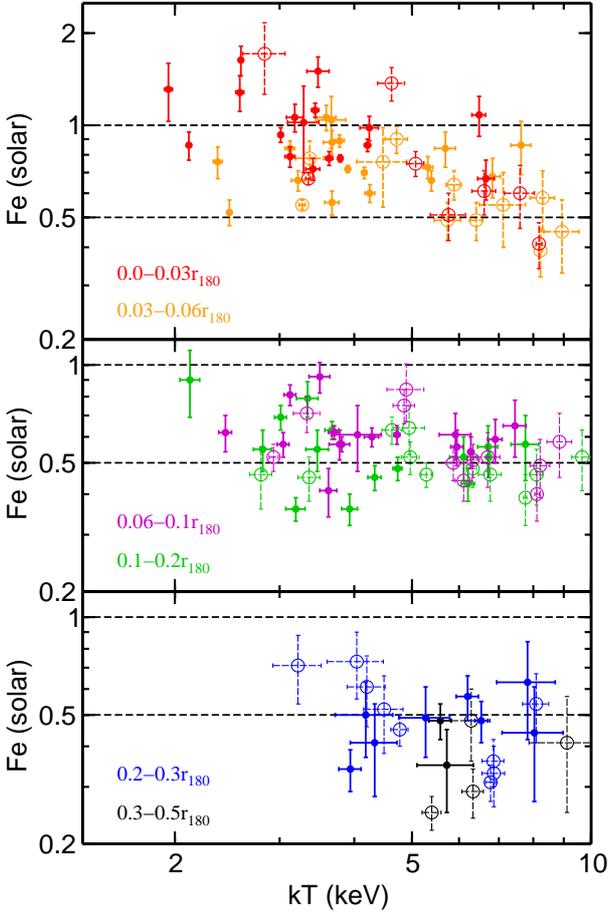}}
\caption{
  Fe abundances derived from  $F_{\rm FeHeK}/F_{3.5-6}$
using the 1T APEC model, plotted against the local ICM temperature. 
The indications of symbols and colors are the same as in Figure
 \ref{feratio}.
}
\label{feabund}
\end{figure}

In Figure \ref{feabund}
the derived Fe abundances are plotted against the ICM temperature.
There is no significant dependence on the plasma  temperature,
or between the cD and non-cD clusters.
\color{black}
The scatter in the Fe abundance tends  to { \color{black} be} larger in lower temperature clusters.
When   { \color{black} the} ICM temperature is higher than 5 keV,
the scatter in the absolute values  { \color{black} of} the Fe abundances at 0.06--0.3$r_{180}$
is  small.

\color{black}

\subsection{Systematic uncertainties in  Fe abundance}
\begin{figure}[tbh]
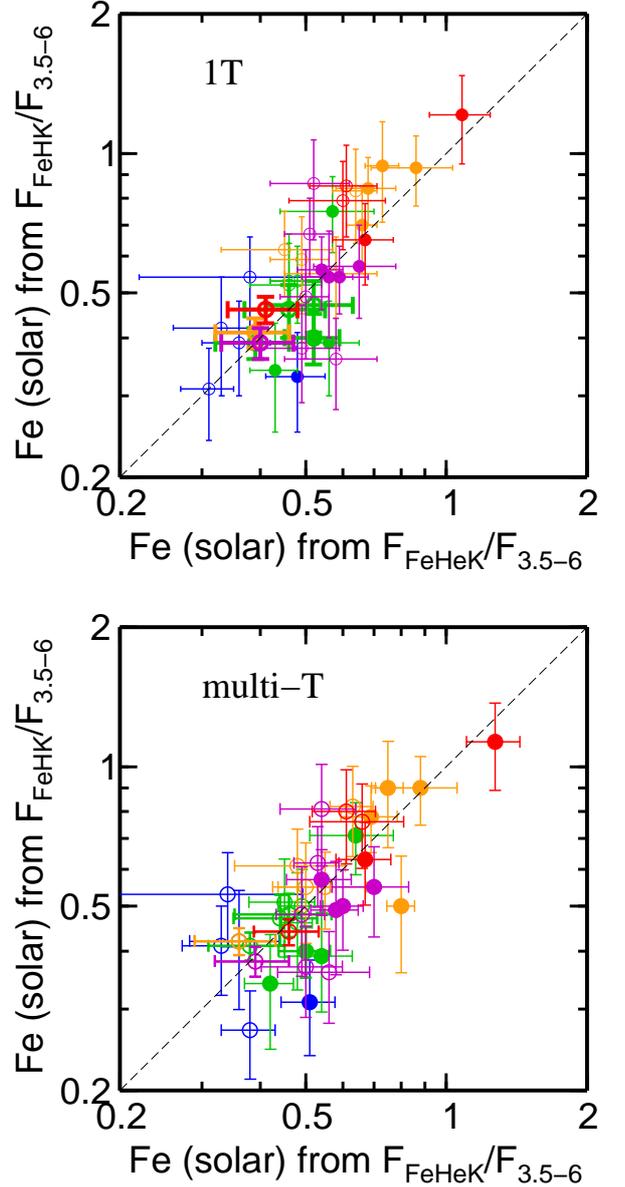

\resizebox{8.1cm}{!}{\includegraphics{13432fig8a.ps}}
\resizebox{8.1cm}{!}{\includegraphics{13432fig8b.ps}}
\caption{
 Fe abundances derived from $F_{\rm FeHK}/F_{3.5-6}$
versus those from  $F_{\rm FeHeK}/F_{3.5-6}$,
\color{black}
using the 1T model (upper panel) and the best-fit multi-T model (lower panel).
A  systematic uncertainty of 10\% in the temperature is considered.
\color{black}
The indications  of colors are the same as in Figure 4.
Closed and open circles correspond to  cD and  non-cD
 clusters, respectively.
}
\label{Fehk}
\end{figure}
\color{black}

In determining the abundance, we must examine uncertainties such as
the  dependence on   the temperature structure  and plasma codes.

\setcounter{table}{2}
\begin{table}
\caption{\color{black} Averages of the relative differences in the derived Fe abundances
[$(\rm{Fe_2-Fe_1})/{\rm Fe_1}$] and their root-mean-square scatters.
Cool (hot)  samples have  local ICM temperatures below (above) 5 keV.
 \color{black}}
\label{tab:dfe}
 \begin{tabular}[t]{cccccc}
\hline\hline
Fe$_1$ & Fe$_2$ & sample & average & scatter \\
\hline
FeHeK, 1T &FeHeK, multi-T & cool &0.01 &0.06 \\
FeHeK, 1T &FeHeK, multi-T & hot &0.02&0.10\\
FeHK, 1T & FeHK, multi-T &hot   &-0.04 & 0.07\\%\hline
\hline
 \end{tabular}
\end{table}
\begin{figure}[h]
 \resizebox{8.1cm}{!}{\includegraphics{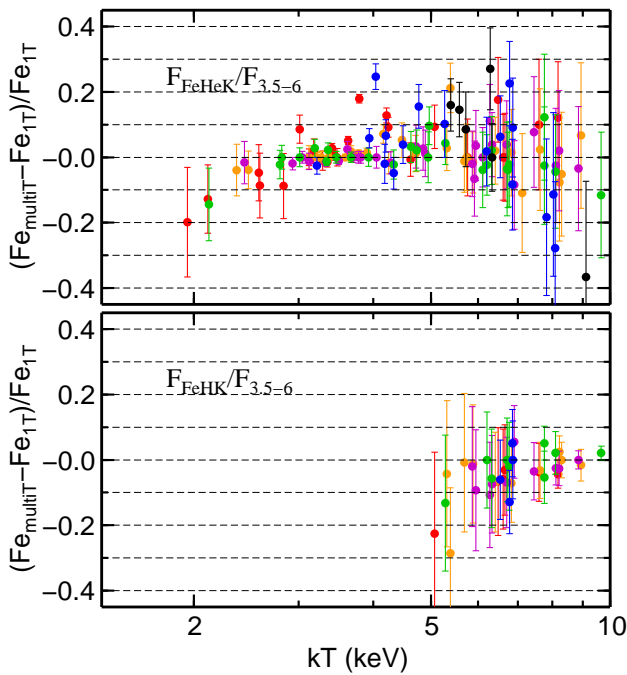}}
\caption
{\color{black}
Differences between the derived  Fe abundances  
using  the best-fit multi-T model and  those using  the 1T model,
\color{black}
plotted against the ICM temperature.
Error bars { \color{black}are} the result of  uncertainties in the ICM temperature
derived from the 1T model, including systematic uncertainties of  10\%.
Upper and lower panels correspond to Fe abundances
derived from $F_{\rm FeHeK}/F_{3.5-6}$ and $F_{\rm FeHK}/F_{3.5-6}$, respectively.
The indications of the colors are the same as in Figure 4.
\color{black}
}
\label{Fesyserror}
\end{figure}

When the ICM temperature derived from the 1T model
was higher than 5 keV, the observed $F_{\rm FeHK}/F_{3.5-6}$ was converted
to the Fe abundance  as described in Section 4.2.
The derived Fe abundances  agree well, within statistical errors,
  with those derived from $F_{\rm FeHeK}/F_{3.5-6}$ (upper panel in Figure \ref{Fehk}, Table 2).
\color{black} 
On average, the difference between  the Fe abundances derived from the two lines
is only a few \%.
\color{black}
At the center of the cool core, where the temperature and abundance gradients 
of the ICM are steep,  the Fe abundances derived from the He-like line
{ \color{black} can be expected to } be higher than those from the H-like line.
However, outside the cool core, where the temperature and abundance gradients 
weaken,  it is reasonable that He-like and H-like lines yield the same Fe abundances.
\color{black}
 \color{black}
Because the temperature dependence of the
He-like and H-like Fe lines are completely different, the consistency
of the Fe abundances \color{black} outside the cool cores \color{black} 
indicates a small systematic
uncertainty owing to the uncertainties in the ICM temperature.

To examine the systematic effect of the assumptions about 
the temperature structure, i.e., the 1T model or the multi-T model,
 we  derived the Fe abundance
from the observed $F_{\rm FeHeK}/F_{3.5-6}$ and
 $F_{\rm FeHK}/F_{3.5-6}$ assuming  the best-fit
multi-T model. For both He-like and H-like lines,
the multi-T model yields
mostly  the same Fe abundances as the 1T model, as
 summarized  in  Figure \ref{Fesyserror} 
and Table \ref{fetable}.
\color{black}
In Figure \ref{Fesyserror}, the error bars arise
from a systematic uncertainty of 10\% in the ICM temperature,
because in most  cases these errors are much larger than those from the
uncertainty in the multi-T model.
Statistical errors in the Fe line strengths are not considered here,
because we use the same line strength.
\color{black}
As summarized in Table \ref{tab:dfe},
   the differences in the Fe abundances
from  $F_{\rm FeHeK}/F_{3.5-6}$ are smaller than several percent
particularly  below 5 keV
owing to the weaker temperature dependence.
Above 5 keV, the systematic differences 
are still smaller than $\sim $10\%.
The differences in the Fe abundances from $F_{\rm FeHK}/F_{3.5-6}$
are also about several percent owing to the weaker temperature dependence.

\color{black}
Figure \ref{Fehk} also compares the Fe abundances from the two lines,
assuming the best-fit multi-temperature model.
Here, statistical errors in the line strengths and a 10\% systematic 
uncertainty in the temperature are considered.
In most clusters, the two lines yield the same Fe abundances,
reflecting  that  the  1T and multi-T models yielded the same Fe abundances.

Higher-temperature components may exist in the ICM.
Therefore, we left  the highest temperature of the multi-T model unconstrained
and fitted the spectra from the  PN and MOS.
For  the ICM temperatures from the 1T model below
 5 keV  we obtained almost the same Fe abundances within several percent.
When the ICM temperature is above 5 keV, the Fe abundances using 
the best-fit multi-T model from the PN spectra were unchanged.
However, using the MOS multi-T model, the Fe abundances from the He-like line
increased by a several tens of percent, whereas those from the H-like line remained the same
within 10\%.
This discrepancy arises from a systematic uncertainty of several percent in the response matrices
of the PN and MOS.
Considering the weak temperature dependence of the $F_{\rm FeHK}/F_{3.5-6}$ at
higher temperatures, the Fe abundances from the H-like line are more reliable
 outside the cool core regions.

\color{black}

Rasia et al. (2008) studied bias or systematic effects in the X-ray spectra using the
spectra simulated by the X-ray Map Simulator and found that
the Fe abundance is recovered with high accuracy for hot ($T>$ 3 keV) and cold
($T<$ 2 keV) systems.
At
 intermediate temperatures (2 keV$< T <$ 3 keV), \color{black} they found  systematic overestimates.
The problem occurs because of  a transition between the relative importance of
the Fe-L and Fe-K lines.
Because we here derived the Fe abundances  from the Fe-K lines,
the observed values do not include those  of the coolest components
(below  $\sim$ 1.7 keV).
The fraction of the emission measure for  temperature components below this temperature is 
smaller than several percent except in 
the innermost region of the coolest cool cores of the Virgo cluster and A~262, 
where the fractions are 30\%.

\color{black}

To examine the systematic effect of the plasma code on
abundance determination, 
we  plotted the temperature dependence of 
the  ratio of line strength to $F_{3.5-6}$ in Figure \ref{feratio} using the MEKAL model 
(Mewe et al. 1995, 1996; Kaastra 1992; Liedahl et al. 1995).
The difference between the MEKAL and APEC models is less than 10\%.

\color{black}
Non-solar abundance ratios of He, C, N, and  O  to  Fe cause a
bias in the derived Fe abundance, since these elements  change
the continuum level.
The difference in the He abundance between the solar abundance tables
by Lodders (2003) and Anders \& Grevesse (1989) yields  a
 bias of several percent.
The bias resulting from a possible high He-abundance in the ICM is discussed in detail in
Ettori \& Fabian (2006) and B{\"o}hringer \& Werner (2010).

\color{black}

In summary, in most  cases, 
the systematic uncertainty in the derived Fe abundances
caused by the  uncertainties in the temperature structures and atomic data
is  about 10\%.

\color{black}

\subsection{Weighted average of  Fe abundance profiles}

\color{black}
We calculated the weighted average of the Fe abundances
derived from  $F_{\rm
 FeHeK}/F_{3.5-6}$ and $F_{\rm FeHK}/F_{3.5-6}$, using the 1T APEC  model
 within the same radial region in the units of $r_{180}$.
 We divided the clusters except for A426 (the Perseus cluster) and A1656 (the Coma cluster) 
according to whether they are  cD  or non-cD clusters.

\begin{figure}[]
\resizebox{8.1cm}{!}{\includegraphics{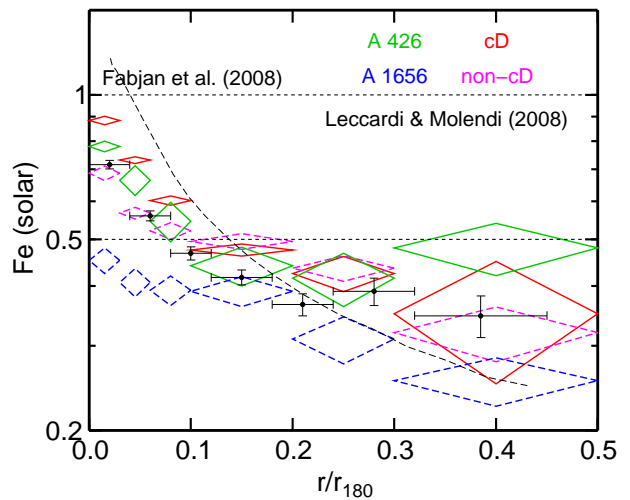}}
\caption{\color{black}  Fe abundance profiles of A426 (green),
A1656 (blue),  and the averages of those of   cD clusters \color{black}
({red} solid diamonds) and  non-cD clusters ({magenta} dashed diamonds) in our sample,
 compared with the average profile of z=0.1$\sim$0.3 clusters ({black} closed circles)
 from Leccardi \& Molendi (2008). {Black} dashed line corresponds to the Fe
 abundance derived from  the numerical simulations { \color{black}
without AGN feedback} by Fabjan et al. (2008).}
\label{fig:feavemolendi}
\end{figure}

The results are summarized in Table \ref{Feave} and Figure \ref{fig:feavemolendi}.
The radial profile of the average Fe abundance of the  cD clusters 
becomes flatter from 0.1 to 0.3 $r_{180}$ at 0.4--0.5 solar.
The Perseus cluster has a 10\% smaller radial profile than   the other cD clusters
within 0.3$r_{180}$, and a flatter Fe abundance profile up to 0.5$r_{180}$.
In contrast, the   non-cD clusters have systematically
lower Fe abundances within 0.1$r_{180}$, and the values
similar to those of cD clusters beyond this radius.
 The Coma cluster also has a flatter radial profile within 0.2$r_{180}$, but at a lower value of
 $\sim$ 0.4 solar,  and beyond this radius,
the Fe abundance gradually decreases to $\sim 0.3$ solar.

\color{black}

\setcounter{table}{3}
\begin{table}[thb]
\caption{Weighted average of Fe abundances derived from $F_{\rm
 FeHeK}/F_{3.5-6}$ and $F_{\rm FeHK}/F_{3.5-6}$,  in the units of the solar
 abundance. 
%Cool and hot clusters have an average ICM temperature lower than 5 keV,
%and higher than 5 keV, respectively, 
}
\label{Feave}
 \begin{tabular}[t]{lcccccccc}
\hline \hline
%        &\multicolumn{3}{c}{cD clusters} & \multicolumn{3}{c}{non-cD clusters} \\
 radius$^a$ &    cDs$^b$ & A~ 426 & non-cDs$^c$ & A~1656\\    
\hline
0.00-0.03 &  0.88$\pm$0.02 &  0.78$\pm$0.01 &  0.69$\pm$0.02&  0.45$\pm$0.03\\
0.03-0.06 &  0.73$\pm$0.01 &  0.66$\pm$0.05 &  0.57$\pm$0.02&  0.41$\pm$0.03\\
0.06-0.10 &  0.60$\pm$0.01 &  0.55$\pm$0.05 &  0.52$\pm$0.02&  0.39$\pm$0.03\\
0.10-0.20 &  0.48$\pm$0.01 &  0.44$\pm$0.04 &  0.50$\pm$0.02&  0.39$\pm$0.03\\
0.20-0.30 &  0.42$\pm$0.04 &  0.41$\pm$0.05 &  0.44$\pm$0.03&  0.31$\pm$0.03\\
0.30-0.50 &  0.35$\pm$0.10 &  0.48$\pm$0.06 &  0.32$\pm$0.04&  0.25$\pm$0.03\\
\hline
\end{tabular}
$^a$: in unit of $r_{180}$.\\
$^b$: the cD clusters except A~426\\
$^c$: the non-cD clusters except A~1656\\
 \end{table}

\subsection{Comparison with previous papers and redshift evolution}
In Figure \ref{fig:feavemolendi} we compare 
our results with those obtained by Leccardi \& Molendi (2008).
Here, the Fe abundances were scaled using Lodders (2003),
\color{black}considering the difference in the solar values of He and Fe. 
\color{black}
Leccardi and Molendi (2008)
 analyzed $\sim$ 50 hot ($kT>3.3$ keV) intermediate-redshift
($z=0.1\sim 0.3$) clusters of galaxies observed with XMM-Newton.
Although we use samples different  from those employed by
Leccardi \& Molendi (2008), 
\color{black}
our
mean profiles of the cD clusters and those of
Leccardi \& Molendi (2008) have remarkably  similar slopes,
but the absolute values differ by 10$\sim$20\%.
\color{black}

Maughan et al. (2008)  analyzed redshift evolution up to
$z\sim 1$ of the Fe
abundance in the ICM using Chandra observations.
The mean value of Maughan et al. (2008)
 at 0.15--1$r_{500}$ of $z<0.5$ clusters is  $\sim 0.5 $ solar using the
solar abundance by Lodders (2003), and 
  is consistent with the average of the cD clusters.

{ \color{black}
These agreements indicate that the evolution of the Fe abundance with redshift
from $z=0$ to $z=0.5$ is consistent with no evolution
 within  our systematic uncertainties.
The expected evolution according to Ettori (2005),
 which is about 30\% from z$\sim 0.5$ to
z$\sim $0, tends to be larger than  our observations.
}

\section{Discussion}

\subsection{\color{black}Flatter radial Fe abundance profile beyond 0.1$r_{180}$}%Enrichment history of Fe in \color{black} {cD}  clusters of galaxie}

\color{black}
Beyond 0.1$r_{180}$, \color{black} well outside the cool cores\color{black},
 both the cD and non-cD clusters have similar Fe
abundance profiles with a relatively small scatter.
\color{black}
Within  0.1--0.3$r_{180}$, the Fe abundances of the sample clusters are 
 0.4$\sim$0.5 solar. \color{black}
There is no dependence on the ICM temperature, i.e., system richness.
In other words, 
outside the central regions of clusters, 
the Fe abundance in the ICM may be universal in clusters of galaxies, and
 clusters of galaxies may have universal metal-enrichment histories.
Beyond 0.3$r_{180}$, the Fe abundances become 0.2--0.5 solar.
\color{black}
These profiles  smoothly connect with the
  Suzaku observations of the Fe distributions in the ICM beyond 0.5$r_{180}$
(Fujita et al. 2008; Tawa 2008).

The distribution of
Fe abundance in the ICM has been modeled via hydrodynamic simulations
(Fabjan et al. 2008, Kapferer et al. 2007).
Figure \ref{fig:feavemolendi} also compares our mean Fe abundance
profiles  with that derived by Fabjan et al. (2008).
Beyond 0.1$r_{180}$, the profile simulated by Fabjan et al. (2008)
continues to decrease  outward, whereas the observed profile
becomes flatter.

\color{black}
The iron-mass-to-light ratio (IMLR), i.e.,
the  ratio of Fe mass in the ICM 
to total stellar luminosity, is important for  studying the origin of Fe,
because Fe is synthesized in stars.
The integrated values of the IMLR profiles of the $kT\sim 3$ keV clusters,
Abell 262,  and AWM7, observed with Suzaku (Sato et al.
2008, 2009), 
increase  outward from \color{black}0.1$r_{180}$ \color{black} to 0.3--0.4$r_{180}$.
This result indicates that Fe in the ICM  extends farther than stars in
the outer regions at least in these two clusters.
\color{black}

A simple explanation is that
 a significant fraction of Fe 
is synthesized in an early phase of cluster evolution.
If metal enrichment occurs after the formation of the clusters, 
the metal distribution is  expected to follow the stellar distribution,
whereas if it occurs before cluster formation, 
we expect a  flatter distribution.
For example,
simulations by Kapferer et al. (2007)  yielded a flatter Fe profile from SN II.
%The fact that
That 
the measurements  of the  Fe abundance  excluding the central region 
by Maughan et al. (2008) are consistent with no evolution at least up to
z=0.7 also supports the early synthesis of  Fe.
Feedback by  active galactic nuclei (AGN) may  also  change the abundance profile.
With  AGN feedback,  a flatter radial profile beyond 0.1$r_{180}$ 
is also obtained in the simulation by Fabjan et al. (2010).

A426 (the Perseus cluster) shows a completely flat Fe profile
from 0.1$r_{180}$ to 0.5$r_{180}$:
 at 0.3--0.5$r_{180}$, the Fe abundance is higher than in  the other clusters.
Within the annular region of 0.3--0.5$r_{180}$, the observed fields of
view do not cover the entire annular region (Figure 1)
and lack the north and south regions.
Several clusters, including A426,  form the Perseus supercluster,
which is  elongated in an east-west direction (Gregory et al. 1981).
The observed very flat Fe abundance may reflect
the history of cluster formation  from filaments of the large-scale structure
of the Universe.

\subsection{Fe profiles within the cool core regions}

Fe in the gas in the cool core of clusters is a mixture of 
that present in the ICM and  later supplies from their cD galaxies.
The latter contains Fe synthesized by SN Ia and
originating from stars through stellar mass loss, because Fe is
synthesized by both SN Ia and SN II.
\color{black}
According to 
  abundance patterns from \color{black} XMM-Newton observations
of the cool cores (Matsushita et al. 2003, 2007; Werner et al. 2006;
de Plaa et al. 2006; Simionescu et al. 2008, 2009) \color{black} 
and Suzaku observations of the cD galaxies of 
AWM7 and A262 (Sato et al. 2008, 2009),
\color{black}
the number ratio of SNe II to SNe Ia is estimated to be 3$\sim$4, \color{black} and
the Fe in the cluster core was synthesized mostly by SN Ia.
\color{black}

\color{black}
Within  cool core regions,
clusters have negative abundance gradients 
 (e.g. Fukazawa et al. 2000, de Grandi et al. 2004).
Rossetti \& Molendi (2010) and Leccardi, Rossetti, \& Molendi (2010)
{ \color{black} also found central metal abundance excess in low-entropy core systems.}
\color{black}
Within { \color{black}these} regions,
observations indicate that metals ejected from cD galaxies { \color{black} have a } more extensive distribution compared with stars.
Analyzing eight cool-core clusters observed with XMM, David \& Nulsen (2008) found 
that Fe in the ICM 
is more extended than the stellar mass,
and excess Fe within the central 100 kpc
 could have been produced by type Ia supernova \color{black} in cD galaxies over the 
past 3--7 Gyr.
\color{black}
The IMLR values in  AWM7 and A262 observed with Suzaku
 have much steeper positive gradients within the cool cores than those beyond 0.1$r_{180}$
(Sato et al. 2008; 2009).
\color{black}

Processes such as jets
from a central AGN as described below,
 or  the ``sloshing'' of cD galaxies in the cluster's
gravitational potential (e.g. Markevitch et al. 2001),
may eject metals from cD galaxies.
On the basis of 
 2D temperature and metallicity maps of M 87 (Simionescu et al. 2008) and Hydra A 
(Simionescu et al. 2009), uplift of Fe by the AGN was discussed in detail.
David \& Nulsen (2008) found that both  turbulent diffusion of entropy
and dissipation are important heating mechanisms in cluster cores.
Rebusco et al. (2006) compared the observed Fe abundance profiles with the predictions
of a model involving  metal ejection from the brightest galaxy and the subsequent diffusion of metals
by stochastic gas motions.
Roediger et al. (2007) studied  the effect of bubble-induced motions on metallicity
profiles through 3D hydro simulations.
The new simulation by Fabjan et al. (2010)
showed a less steep central Fe peak \color{black} including feedback from AGN
in cD galaxies\color{black}.
The observed  small scatter in the Fe abundance slope within
0.1$r_{180}$ in the cD clusters
 suggests a common process of metal supply and ejection in 
 the cD galaxies.
The next Japanese X-ray satellite, ASTRO-H,  will directly reveal line broadening
caused by turbulence in the cores of clusters.

Peng \& Nagai (2009) pointed out that the He/H abundance ratio 
could  be four times the  solar abundance  at the center of massive clusters.
Such a high He abundance causes an underestimation  of the Fe abundance by a
factor of two (Ettori \& Fabian 2006).
However,  in actual clusters, metals at the center of the cluster come from
mass loss by central galaxies, and in addition, more extensively distributed compared with stars.
Therefore, the centers of clusters should not be static,  as assumed in 
the calculations of Peng \& Nagai (2009).

\color{black}

\subsection{Mixing of the  ICM in  non-cD clusters}
\color{black}

Most of 
clusters without cool cores tend to show
 no central abundance excess as found in
 Fukazawa et al. (2000); de Grandi et al. (2004).
 Leccardi, Rossatti \& Molendi (2010) found  that 
{ \color{black} some medium- and high-entropy core systems
have central abundance excess, most of which 
 are undergoing a phase of rapid dynamical changes,
and suggested that these clusters originate from {\color{black}low-entropy core} systems.}

Our results are consistent with the discussion in Leccardi, 
Rossatti \& Molendi (2010).
The non-cD clusters in this paper include merging clusters in various stages,
from  heavily distorted morphologies
to relatively relaxed ones without cool cores.
During cluster merging, mixing of the ICM { \color{black} could} destroy
 the central Fe peak.
Clusters with the highest central Fe abundances have highly disturbed
morphologies.
In the first stage of  merging,
the Fe peak and the cool core may remain.
Then,  mixing of the ICM destroys the Fe peak, 
and the Fe abundance  becomes flat at the same
level as in   cD clusters at 0.1--0.5$r_{180}$.

\color{black}
\section{Summary and conclusion}

We derived radial profiles of the
Fe abundance of the ICM in nearby ($z<0.08$) clusters
observed with XMM-Newton.
The Fe abundances of the ICM were derived from the ratio of the flux 
  of the K$\alpha$ lines of  He-like or H-like Fe  
 to those of  the continuum  at 3.5--6.0 keV,
because the systematic uncertainty in the  continuum flux in this energy 
band owing to the background  is  smaller.
The temperature dependence of these ratios
 constrains the  Fe abundances of multi-temperature plasmas.

\color{black}
In cluster core regions ($<0.1 r_{180}$), 
the observed Fe abundances of cD clusters show similar radial profiles.
The less-peaked abundance profile \color{black} compared with the light \color{black} in the central region indicates the
 ejection of  metals from cD galaxies.

\color{black}
In the outer regions, 0.1--0.2$r_{180}$, Fe abundances of 0.4--0.5 solar
appear  to be universal
 with no temperature dependence.
The observed flatter Fe abundance profiles of the cD clusters
beyond  0.1$r_{180}$ indicate
early metal enrichment.

Chandra and XMM cannot reach beyond $0.5 r_{180}$ owing to a 
 high,   unstable particle background.  Information is 
available for only about 10\% of the cluster volume,
 and the majority has not yet been revealed.
The total amount of Fe synthesized in galaxies can be 
derived only by the precise abundance measurements beyond $0.5 r_{180}$.
Because of  its low background,   Suzaku is the only satellite available
for the next several years to study clusters of galaxies up to the virial
radius. With Suzaku, Fe abundances will be determined out to a radius of
0.7$\sim$ 0.8$r_{180}$.

\longtab{2}{
\begin{longtable}{lllllc}
\caption{\label{fetable}Results of spectral fitting of annular spectra.}\\
%\label{fetable}
\hline \hline
 cluster		  & r ($r_{180}$)&$kT^a$ (keV) & Fe (solar) from  $F_{\rm
 FeHeK}/F_{3.5-6}$ & Fe (solar) from $F_{\rm FeHK}/F_{3.5-6}$\\
% \\ & &  &   $F_{\rm FeHeK}/F_{3.5-6}$ &$F_{\rm FeHK}/F_{3.5-6}$\\
  &   & %Best-fit$^a\pm\Delta_{\rm stat}^b\Delta_{\rm sys}^c$  
&
  Best-fit$^b\pm\Delta_{\rm stat}^c\Delta_{\rm sys}^d$ &  Best-fit$^b\pm\Delta_{\rm stat}^c\Delta_{\rm sys}^d$ \\ \hline
\endfirsthead
\hline \hline
 cluster	  & r ($r_{180}$)&$kT^a$ (keV) & Fe (solar) from  $F_{\rm
 FeHeK}/F_{3.5-6}$ & Fe (solar) from $F_{\rm FeHK}/F_{3.5-6}$\\
% \\ & &  &   $F_{\rm FeHeK}/F_{3.5-6}$ &$F_{\rm FeHK}/F_{3.5-6}$\\
  &   & &
  Best-fit$^b\pm\Delta_{\rm stat}^c\Delta_{\rm sys}^d$ &  Best-fit$^b\pm\Delta_{\rm stat}^c\Delta_{\rm sys}^d$ \\ \hline
\endhead
%\endfoot
  \hline
\multicolumn{5}{l}{$^a$Best-fit ICM temperature with 1$\sigma$
statistical error derived from the
1T model fits}\\
%\multicolumn{5}{l}{$^b$1 $\sigma $ statistical error }\\
%\multicolumn{5}{l}{$^c$ The difference in the best-fit temperatures
%from 1T model fits without and with the ``powerlaw/b'' component.}\\
\multicolumn{5}{l}{$^b$Best-fit Fe abundance  using the ICM
 temperature derived from the 1T model fits}\\
%\multicolumn{5}{l}{with the ``powerlaw/b'' component}\\
\multicolumn{5}{l}{$^c$1 $\sigma $ statistical error, considering a 10 \%
systematic uncertainty in the ICM temperature 
 derived from the 1T model.}\\
%\multicolumn{5}{l}{$^f$ The difference in the derived Fe abundances
%from the 1T model fits without and with the ``powerlaw/b'' component.}\\
\multicolumn{5}{l}{$^d$ Difference in the derived Fe abundances
using the best-fit multi-T model and the 1T model}\\
%\multicolumn{5}{l}{ ``powerlaw/b'' component}\\
%
\endlastfoot
%\begin{tabular}[t]
\hline
A262 &0.00-0.03 & 1.95$\pm$0.03 & 1.31$\pm$0.28-0.26 & ---&     \\
A262 &0.03-0.06 & 2.36$\pm$0.04 & 0.76$\pm$0.09-0.03 & ---&     \\
A262 &0.06-0.10 & 2.43$\pm$0.06 & 0.62$\pm$0.08-0.02 & ---&     \\
A262 &0.10-0.20 & 2.12$\pm$0.08 & 0.90$\pm$0.21-0.13 & ---&     \\
Virgo &0.00-0.03 & 2.11$\pm$0.00 & 0.86$\pm$0.09-0.11 & ---&     \\
Virgo &0.03-0.06 & 2.47$\pm$0.01 & 0.52$\pm$0.05-0.02 & ---&     \\
A4038 &0.00-0.03 & 3.12$\pm$0.06 & 0.79$\pm$0.06+0.00 & ---&     \\
A4038 &0.03-0.06 & 3.22$\pm$0.08 & 0.66$\pm$0.05+0.00 & ---&     \\
A4038 &0.06-0.10 & 3.04$\pm$0.07 & 0.57$\pm$0.05+0.00 & ---&     \\
A4038 &0.10-0.20 & 2.81$\pm$0.10 & 0.55$\pm$0.08+0.01 & ---&     \\
A1060 &0.00-0.03 & 3.35$\pm$0.03 & 0.67$\pm$0.03-0.01 & ---&     \\
A1060 &0.03-0.06 & 3.27$\pm$0.04 & 0.55$\pm$0.02-0.00 & ---&     \\
A1060 &0.06-0.10 & 2.93$\pm$0.05 & 0.52$\pm$0.05-0.01 & ---&     \\
A1060 &0.10-0.15 & 2.79$\pm$0.12 & 0.46$\pm$0.10-0.01 & ---&     \\
A2052 &0.00-0.03 & 2.57$\pm$0.04 & 1.28$\pm$0.17-0.07 & ---&     \\
A2052 &0.03-0.06 & 3.12$\pm$0.06 & 0.84$\pm$0.05-0.00 & ---&     \\
A2052 &0.06-0.10 & 3.12$\pm$0.07 & 0.81$\pm$0.06-0.00 & ---&     \\
A2052 &0.10-0.20 & 3.01$\pm$0.07 & 0.69$\pm$0.06-0.01 & ---&     \\
A1367 &0.00-0.03 & 2.83$\pm$0.23 & 1.71$\pm$0.45-0.15 & ---&     \\
A1367 &0.03-0.06 & 3.37$\pm$0.18 & 0.78$\pm$0.11+0.00 & ---&     \\
A1367 &0.06-0.10 & 3.33$\pm$0.17 & 0.71$\pm$0.09-0.01 & ---&     \\
A1367 &0.10-0.20 & 3.36$\pm$0.14 & 0.45$\pm$0.07+0.00 & ---&     \\
A1367 &0.20-0.30 & 3.22$\pm$0.30 & 0.71$\pm$0.17-0.02 & ---&     \\
A780 &0.00-0.03 & 3.41$\pm$0.09 & 0.72$\pm$0.06+0.02 & ---&     \\
A780 &0.03-0.06 & 3.67$\pm$0.10 & 0.56$\pm$0.05+0.01 & ---&     \\
A780 &0.06-0.10 & 3.62$\pm$0.12 & 0.41$\pm$0.07+0.01 & ---&     \\
A780 &0.10-0.20 & 3.93$\pm$0.12 & 0.36$\pm$0.04+0.00 & ---&     \\
A780 &0.20-0.30 & 4.33$\pm$0.39 & 0.41$\pm$0.13-0.02 & ---&     \\
A2589 &0.00-0.03 & 3.48$\pm$0.15 & 1.50$\pm$0.17-0.01 & ---&     \\
A2589 &0.03-0.06 & 3.59$\pm$0.14 & 1.06$\pm$0.10-0.01 & ---&     \\
A2589 &0.06-0.10 & 3.50$\pm$0.14 & 0.92$\pm$0.10-0.01 & ---&     \\
A2589 &0.10-0.20 & 3.34$\pm$0.14 & 0.79$\pm$0.10-0.01 & ---&     \\
AWM7 &0.00-0.03 & 3.44$\pm$0.05 & 1.12$\pm$0.06+0.01 & ---&     \\
AWM7 &0.03-0.06 & 3.78$\pm$0.07 & 0.89$\pm$0.04+0.00 & ---&     \\
AWM7 &0.06-0.10 & 3.70$\pm$0.08 & 0.62$\pm$0.03+0.00 & ---&     \\
AWM7 &0.10-0.19 & 3.67$\pm$0.10 & 0.63$\pm$0.04+0.00 & ---&     \\
MKW3s &0.00-0.03 & 3.18$\pm$0.10 & 1.06$\pm$0.11+0.01 & ---&     \\
MKW3s &0.03-0.06 & 3.67$\pm$0.12 & 0.88$\pm$0.08+0.00 & ---&     \\
MKW3s &0.06-0.10 & 3.78$\pm$0.15 & 0.57$\pm$0.06+0.00 & ---&     \\
MKW3s &0.10-0.20 & 3.47$\pm$0.15 & 0.55$\pm$0.08+0.01 & ---&     \\
A2063 &0.00-0.03 & 3.29$\pm$0.20 & 1.02$\pm$0.32-0.01 & ---&     \\
A2063 &0.03-0.06 & 3.66$\pm$0.21 & 1.04$\pm$0.20+0.01 & ---&     \\
A2063 &0.06-0.10 & 4.05$\pm$0.39 & 0.61$\pm$0.14-0.01 & ---&     \\
A3526 &0.00-0.03 & 2.58$\pm$0.01 & 1.63$\pm$0.18-0.14 & ---&     \\
A3526 &0.03-0.06 & 3.90$\pm$0.03 & 0.72$\pm$0.02+0.00 & ---&     \\
A3526 &0.06-0.10 & 3.81$\pm$0.03 & 0.57$\pm$0.03+0.01 & ---&     \\
A3526 &0.10-0.20 & 3.19$\pm$0.12 & 0.36$\pm$0.03+0.01 & ---&     \\
A2199 &0.00-0.03 & 3.63$\pm$0.05 & 0.78$\pm$0.04+0.04 & ---&     \\
A2199 &0.03-0.06 & 4.25$\pm$0.08 & 0.60$\pm$0.04-0.01 & ---&     \\
A2199 &0.06-0.10 & 4.28$\pm$0.12 & 0.60$\pm$0.04-0.01 & ---&     \\
A2199 &0.10-0.20 & 4.33$\pm$0.11 & 0.45$\pm$0.04-0.01 & ---&     \\
A2199 &0.20-0.30 & 4.18$\pm$0.46 & 0.50$\pm$0.13-0.01 & ---&     \\
A496 &0.00-0.03 & 3.01$\pm$0.02 & 0.93$\pm$0.05+0.08 & ---&     \\
A496 &0.03-0.06 & 4.16$\pm$0.03 & 0.70$\pm$0.03+0.05 & ---&     \\
A496 &0.06-0.10 & 4.72$\pm$0.05 & 0.61$\pm$0.04+0.01 & ---&     \\
A496 &0.10-0.20 & 4.73$\pm$0.08 & 0.48$\pm$0.04+0.01 & ---&     \\
A496 &0.20-0.30 & 3.94$\pm$0.17 & 0.34$\pm$0.05+0.03 & ---&     \\
%A1644 &0.00-0.03 & 2.72$\pm$0.25 & 0.00$\pm$2.65+0.00 & ---&     \\
A1644 &0.03-0.06 & 4.47$\pm$0.43 & 0.76$\pm$0.22+0.04 & ---&     \\
A1644 &0.06-0.10 & 4.89$\pm$0.35 & 0.84$\pm$0.17+0.01 & ---&     \\
A1644 &0.10-0.20 & 4.95$\pm$0.30 & 0.64$\pm$0.11-0.00 & ---&     \\
A1644 &0.20-0.30 & 4.20$\pm$0.31 & 0.61$\pm$0.15+0.04 & ---&     \\
A3562 &0.00-0.03 & 4.62$\pm$0.24 & 1.37$\pm$0.17-0.01 & ---&     \\
A3562 &0.03-0.06 & 4.71$\pm$0.20 & 0.90$\pm$0.09+0.01 & ---&     \\
A3562 &0.06-0.10 & 4.85$\pm$0.20 & 0.75$\pm$0.09+0.02 & ---&     \\
A3562 &0.10-0.20 & 4.63$\pm$0.18 & 0.63$\pm$0.06+0.02 & ---&     \\
A3562 &0.20-0.30 & 4.49$\pm$0.34 & 0.52$\pm$0.14+0.02 & ---&     \\
A3558 &0.00-0.03 & 5.07$\pm$0.17 & 0.75$\pm$0.07+0.07 & ---&     \\
A3558 &0.03-0.06 & 5.88$\pm$0.17 & 0.64$\pm$0.07-0.01 &0.83$\pm$0.19-0.01 &   \\
A3558 &0.06-0.10 & 5.87$\pm$0.16 & 0.50$\pm$0.06-0.01 &0.49$\pm$0.13-0.01 &   \\
A3558 &0.10-0.20 & 5.29$\pm$0.11 & 0.46$\pm$0.04+0.02 &0.53$\pm$0.14-0.07 &   \\
A3558 &0.20-0.30 & 4.77$\pm$0.16 & 0.45$\pm$0.05+0.07 & ---&     \\
A3627 &0.00-0.03 & 5.76$\pm$0.39 & 0.51$\pm$0.09-0.00 & ---&     \\
A3627 &0.03-0.06 & 5.74$\pm$0.29 & 0.49$\pm$0.07+0.01 & ---&     \\
A3627 &0.06-0.10 & 6.11$\pm$0.20 & 0.44$\pm$0.06-0.00 & ---&     \\
A3627 &0.10-0.20 & 4.97$\pm$0.17 & 0.52$\pm$0.06+0.06 & ---&     \\
A3627 &0.20-0.30 & 4.04$\pm$0.44 & 0.73$\pm$0.17+0.17 & ---&     \\
A1795 &0.00-0.03 & 4.21$\pm$0.05 & 0.86$\pm$0.04+0.11 & ---&     \\
A1795 &0.03-0.06 & 5.32$\pm$0.10 & 0.73$\pm$0.06+0.02 &0.94$\pm$0.23-0.04 &   \\
A1795 &0.06-0.10 & 5.95$\pm$0.14 & 0.56$\pm$0.07+0.02 &0.54$\pm$0.14-0.06 &   \\
A1795 &0.10-0.20 & 6.21$\pm$0.16 & 0.43$\pm$0.05-0.01 &0.34$\pm$0.09+0.00 &   \\
A1795 &0.20-0.30 & 6.20$\pm$0.27 & 0.57$\pm$0.09+0.00 & ---&     \\
A1795 &0.30-0.50 & 5.72$\pm$0.62 & 0.35$\pm$0.10+0.03 & ---&     \\
A85 &0.00-0.03 & 4.24$\pm$0.15 & 0.98$\pm$0.09+0.09 & ---&     \\
A85 &0.03-0.06 & 5.69$\pm$0.23 & 0.84$\pm$0.11-0.01 & ---&     \\
A85 &0.06-0.10 & 5.92$\pm$0.36 & 0.61$\pm$0.10-0.04 & ---&     \\
A85 &0.10-0.20 & 6.11$\pm$0.32 & 0.52$\pm$0.08-0.02 & ---&     \\
A85 &0.20-0.30 & 5.27$\pm$0.52 & 0.49$\pm$0.12+0.05 & ---&     \\
A3667 &0.00-0.03 & 6.61$\pm$0.30 & 0.61$\pm$0.10-0.00 &0.85$\pm$0.19-0.05 &   \\
A3667 &0.03-0.06 & 6.41$\pm$0.16 & 0.49$\pm$0.07+0.01 &0.59$\pm$0.14-0.04 &   \\
A3667 &0.06-0.10 & 6.33$\pm$0.14 & 0.51$\pm$0.06+0.02 &0.67$\pm$0.13-0.05 &   \\
A3667 &0.10-0.20 & 6.32$\pm$0.10 & 0.48$\pm$0.06+0.01 &0.53$\pm$0.10-0.03 &   \\
A3667 &0.20-0.30 & 6.85$\pm$0.29 & 0.36$\pm$0.06-0.04 &0.39$\pm$0.09+0.02 &   \\
A3667 &0.30-0.50 & 6.33$\pm$0.26 & 0.29$\pm$0.05+0.01 & ---&     \\
A426 &0.00-0.03 & 3.79$\pm$0.01 & 0.78$\pm$0.02+0.13 & ---&     \\
A426 &0.03-0.06 & 5.39$\pm$0.04 & 0.66$\pm$0.05+0.14 &0.70$\pm$0.15-0.20 &   \\
A426 &0.06-0.10 & 6.28$\pm$0.07 & 0.54$\pm$0.06+0.06 &0.56$\pm$0.10-0.06 &   \\
A426 &0.10-0.20 & 6.73$\pm$0.07 & 0.52$\pm$0.07-0.01 &0.40$\pm$0.05+0.01 &   \\
A426 &0.20-0.30 & 6.54$\pm$0.16 & 0.48$\pm$0.07+0.03 &0.33$\pm$0.08-0.02 &   \\
A426 &0.30-0.50 & 5.58$\pm$0.24 & 0.48$\pm$0.06+0.07 & ---&     \\
A2256 &0.03-0.06 & 7.12$\pm$0.63 & 0.55$\pm$0.15-0.06 & ---&     \\
A2256 &0.06-0.10 & 6.70$\pm$0.33 & 0.52$\pm$0.10+0.02 &0.86$\pm$0.21-0.05 &   \\
A2256 &0.10-0.20 & 6.77$\pm$0.30 & 0.46$\pm$0.08-0.01 &0.52$\pm$0.12-0.01 &   \\
A2256 &0.20-0.30 & 8.08$\pm$0.41 & 0.54$\pm$0.13-0.15 & ---&     \\
A2256 &0.30-0.50 & 9.11$\pm$1.25 & 0.41$\pm$0.16-0.15 & ---&     \\
A3571 &0.00-0.03 & 6.66$\pm$0.21 & 0.67$\pm$0.10-0.01 &0.65$\pm$0.13-0.02 &   \\
A3571 &0.03-0.06 & 6.83$\pm$0.21 & 0.68$\pm$0.10+0.01 &0.84$\pm$0.14-0.06 &   \\
A3571 &0.06-0.10 & 6.90$\pm$0.19 & 0.59$\pm$0.09-0.05 &0.54$\pm$0.09+0.03 &   \\
A3571 &0.10-0.20 & 6.71$\pm$0.26 & 0.56$\pm$0.09-0.02 &0.39$\pm$0.09-0.01 &   \\
A3571 &0.20-0.30 & 7.82$\pm$0.88 & 0.63$\pm$0.21-0.12 & ---&     \\
A2029 &0.00-0.03 & 6.48$\pm$0.17 & 1.08$\pm$0.16+0.19 &1.21$\pm$0.26-0.08 &   \\
A2029 &0.03-0.06 & 7.63$\pm$0.29 & 0.86$\pm$0.17+0.02 &0.93$\pm$0.16-0.03 &   \\
A2029 &0.06-0.10 & 7.45$\pm$0.33 & 0.65$\pm$0.13+0.05 &0.57$\pm$0.13-0.02 &   \\
A2029 &0.10-0.20 & 7.75$\pm$0.41 & 0.57$\pm$0.13+0.07 &0.75$\pm$0.14-0.03 &   \\
A2029 &0.20-0.30 & 8.03$\pm$0.94 & 0.44$\pm$0.17-0.05 & ---&     \\
A1656 &0.00-0.03 & 8.17$\pm$0.09 & 0.41$\pm$0.07+0.05 &0.46$\pm$0.03-0.02 &   \\
A1656 &0.03-0.06 & 8.22$\pm$0.06 & 0.39$\pm$0.07-0.03 &0.41$\pm$0.03+0.01 &   \\
A1656 &0.06-0.10 & 8.11$\pm$0.07 & 0.40$\pm$0.07-0.00 &0.39$\pm$0.03-0.00 &   \\
A1656 &0.10-0.20 & 7.75$\pm$0.05 & 0.39$\pm$0.07-0.02 &0.39$\pm$0.03+0.01 &   \\
A1656 &0.20-0.30 & 6.78$\pm$0.13 & 0.31$\pm$0.04+0.08 &0.31$\pm$0.07-0.04 &   \\
A1656 &0.30-0.50 & 5.40$\pm$0.20 & 0.25$\pm$0.03+0.03 & ---&     \\
A3266 &0.00-0.03 & 7.59$\pm$0.44 & 0.60$\pm$0.14+0.06 &0.79$\pm$0.17-0.04 &   \\
A3266 &0.03-0.06 & 8.29$\pm$0.39 & 0.58$\pm$0.13-0.02 &0.55$\pm$0.11-0.00 &   \\
A3266 &0.06-0.10 & 8.21$\pm$0.31 & 0.49$\pm$0.10+0.00 &0.38$\pm$0.09-0.01 &   \\
A3266 &0.10-0.20 & 8.09$\pm$0.21 & 0.46$\pm$0.09-0.02 &0.46$\pm$0.06+0.01 &   \\
A3266 &0.20-0.30 & 6.86$\pm$0.30 & 0.33$\pm$0.07+0.03 &0.42$\pm$0.12-0.00 &   \\
A3266 &0.30-0.50 & 6.29$\pm$0.47 & 0.48$\pm$0.12+0.13 & ---&     \\
A754 &0.03-0.06 & 8.93$\pm$0.60 & 0.45$\pm$0.12+0.03 &0.62$\pm$0.13-0.01 &   \\
A754 &0.06-0.10 & 8.85$\pm$0.42 & 0.58$\pm$0.13-0.02 &0.36$\pm$0.08+0.01 &   \\
A754 &0.10-0.20 & 9.66$\pm$0.38 & 0.52$\pm$0.11-0.06 &0.47$\pm$0.06+0.01 &   \\
A754 &0.20-0.30 & 10.04$\pm$1.67 & 0.38$\pm$0.16-0.04 &0.54$\pm$0.12-0.01 &   \\

%\input plot7/Fe_20090912_i4.tex
%		 \end{tabular}
 \end{longtable}
}


\begin{thebibliography}{DUM}




\bibitem[Anders \& Grevesse(1989)]{angr1989}
 Anders, E., \& Grevesse, N.\ 1989, \gca, 53, 197 

\bibitem[Arnaud et al.(1992)]{arnaud1992} Arnaud, M., Rothenflug, 
R., Boulade, O., Vigroux, L., \& Vangioni-Flam, E.\ 1992, \aap, 254, 49 

\bibitem[Baldi et al.(2007)]{2007ApJ...666..835B} Baldi, A., Ettori, S., 
Mazzotta, P., Tozzi, P., \& Borgani, S.\ 2007, \apj, 666, 835 

\bibitem{} B\"ohringer, H., Matsushita, K., Churazov, E. et al. 2002,
	A\&A, 382,804
%\bibitem[B{\"o}hringer 
%\& Werner(2010)]{2010A&ARv..18..127B} B{\"o}hringer, H., \& Werner, N.\ 2010, \aapr, 18, 127 




\bibitem[loddbias]{2010A&ARv..18..127B} B{\"o}hringer, H., \& Werner, N.\ 2010, \aapr, 18, 127 
%\bibitem[Croston et al. 2008]{Croston2008} Croston, J.~H., et al.\ 2008, \aap, 487, 431 
\bibitem[David 
\& Nulsen(2008)]{2008ApJ...689..837D} David, L.~P., \& Nulsen, P.~E.~J.\ 2008, \apj, 689, 837 



\bibitem[De Grandi et al.(2004)]{2004A&A...419....7D} De Grandi, S., Ettori, S., Longhetti, M., \& Molendi, S.\ 2004, \aap, 419, 7 
\bibitem[de Plaa et 
al.(2006)]{2006A&A...452..397D} de Plaa, J., et al.\ 2006, \aap, 452, 397 


\bibitem[Dickey 
\& Lockman(1990)]{1990ARA&A..28..215D} Dickey, J.~M., \& Lockman, F.~J.\ 1990, \araa, 28, 215 



\bibitem[Ebeling et al.(1996)]{1996MNRAS.281..799E} Ebeling, H., Voges, W., 
Bohringer, H., Edge, A.~C., Huchra, J.~P., 
\& Briel, U.~G.\ 1996, \mnras, 281, 799 
\bibitem[Ettori(2005)]{2005MNRAS.362..110E} Ettori, S.\ 2005, \mnras, 362, 
110 
\bibitem[Ettori 
\& Fabian(2006)]{2006MNRAS.369L..42E} Ettori, S., \& Fabian, A.~C.\ 2006, \mnras, 369, L42 



\bibitem[Ezawa et al.(1997)]{1997ApJ...490L..33E} Ezawa, H., Fukazawa, Y., 
Makishima, K., Ohashi, T., Takahara, F., Xu, H., 
\& Yamasaki, N.~Y.\ 1997, \apjl, 490, L33 
\bibitem[Ezawa et al.(2001)]{2001PASJ...53..595E} Ezawa, H., et al.\ 2001, 
\pasj, 53, 595 




\bibitem[Fabjan et al.(2008)]{2008MNRAS.386.1265F} Fabjan, D., Tornatore, 
L., Borgani, S., Saro, A., \& Dolag, K.\ 2008, \mnras, 386, 1265
\bibitem[Fabjan et al.(2010)]{2010MNRAS.401.1670F} Fabjan, D., Borgani, S., 
Tornatore, L., Saro, A., Murante, G., \& Dolag, K.\ 2010, \mnras, 401, 1670 

\bibitem[Finoguenov et al.(2000)]{2000ApJ...544..188F} Finoguenov, A., 
David, L.~P., \& Ponman, T.~J.\ 2000, \apj, 544, 188 
\bibitem[Finoguenov et al.(2001)]{2001ApJ...555..191F} Finoguenov, A., 
Arnaud, M., \& David, L.~P.\ 2001, \apj, 555, 191 

\bibitem[Fujita et al.(2008)]{2008PASJ...60S.343F} Fujita, Y., Tawa, N., 
Hayashida, K., Takizawa, M., Matsumoto, H., Okabe, N., 
\& Reiprich, T.~H.\ 2008, \pasj, 60, 343 


\bibitem[Fukazawa et al.(1998)]{1998PASJ...50..187F} Fukazawa, Y., 
Makishima, K., Tamura, T., Ezawa, H., Xu, H., Ikebe, Y., Kikuchi, K., \& 
Ohashi, T.\ 1998, \pasj, 50, 187 

\bibitem[Fukazawa et al.(2000)]{fukazawa2000} Fukazawa, Y., 
Makishima, K., Tamura, T., Nakazawa, K., Ezawa, H., Ikebe, Y., Kikuchi, K., 
\& Ohashi, T.\ 2000, \mnras, 313, 21 
\bibitem[Gregory et al.(1981)]{1981ApJ...243..411G} Gregory, S.~A., 
Thompson, L.~A., \& Tifft, W.~G.\ 1981, \apj, 243, 411 


\bibitem{}Kaastra ~J.S. 1992, An X-Ray Spectral Code for Optically Thin Plasmas (Internal SRON-Leiden Report, updated version 2.0)
\bibitem{} Katayama, H., Takahashi, I., Ikebe, Y., Matsushita, K.,
        \& Freyberg, M., 2004, A\&A, 414, 767
\bibitem[Kapferer et 
al.(2007)]{2007A&A...466..813K} Kapferer, W., et al.\ 2007, \aap, 466, 813 

\bibitem[Leccardi 
\&  Molendi(2008)]{molendi2008} Leccardi, A., \& Molendi, S.\ 2008, \aap, 487, 461 
\bibitem[Leccardi et 
al.(2010)]{2010A&A...510A..82L} Leccardi, A., Rossetti, M., \& Molendi, S.\ 2010, \aap, 510, A82 



\bibitem{}Liedahl ~D. A., Osterheld ~A.L., \& Goldstein ~W.H.  1995, ApJ, 438,     L115

%\bibitem[Lodders et al.(2003)]{lodd2003}
% Lodders, K. et al.\ 2003, \apj, 591, 1220
\bibitem[Lodders(2003)]{2003ApJ...591.1220L} Lodders, K.\ 2003, \apj, 591, 
1220 


\bibitem[Lumb et al.(2002)]{Lumb2002} Lumb, D.~H., Warwick, 
R.~S., Page, M., \& De Luca, A.\ 2002, \aap, 389, 93 
\bibitem[Matsushita et  al.(2003)]{Matsushita2003} Matsushita, K., Finoguenov, A., B\"ohringer, H.\ 2003, \aap, 401, 443 


\bibitem[Matsushita et al.(2007)]{Matsushita2007b} % Centaurus Newton
 Matsushita, K., B{\"o}hringer, H., Takahashi, I., \& Ikebe, Y.\
 2007, \aap, 462, 953 

\bibitem{} Mewe ~R., Gronenschild ~E.H.B.M., \& van~den~Oord ~G.H.J. 1985, A\&AS, 62,197
\bibitem{} Mewe ~R., Lemen ~J.R., \& van~den~Oord,~G.H.J. 1986, A\&AS,
        65,511

\bibitem[Makishima et al.(2001)]{Makishima2001}
 Makishima, K., Ezawa, H., Fukuzawa, Y., Honda, H., Ikebe, Y., Kamae, T., 
 Kikuchi, K., Matsushita, K., Nakazawa, K., Ohashi, T., Takahashi, T., 
 Tamura, T. \& Xu, H.\ 2001, \pasj, 53, 401 

\bibitem[Markevitch et al.(1998)]{Markevitch1998}
 Markevitch, M., Forman, W.~R., Sarazin, C.~L., \& Vikhlinin, A.\
 1998, \apj, 503, 77 

\bibitem[Markevitch et al.(2001)]{Maxim2001} Markevitch, M., 
Vikhlinin, A., \& Mazzotta, P.\ 2001, \apjl, 562, L153 
\bibitem[Maughan et al.(2008)]{maughan2008} Maughan, B.~J., Jones, 
C., Forman, W., \& Van Speybroeck, L.\ 2008, \apjs, 174, 117 
\bibitem[Peng 
\& Nagai(2009)]{2009ApJ...693..839P} Peng, F., \& Nagai, D.\ 2009, \apj, 693, 839 



\bibitem[Rasia et al.(2008)]{2008ApJ...674..728R} Rasia, E., Mazzotta, P., 
Bourdin, H., Borgani, S., Tornatore, L., Ettori, S., Dolag, K., 
\& Moscardini, L.\ 2008, \apj, 674, 728 

\bibitem[Rebusco et al.(2006)]{2006MNRAS.372.1840R} Rebusco, P., Churazov, 
E., B{\"o}hringer, H., \& Forman, W.\ 2006, \mnras, 372, 1840 
%\bibitem[Reese et al.(2010)]{2010arXiv1006.4486R} Reese, E.~D., Kawahara, 
%H., Kitayama, T., Ota, N., Sasaki, S., \& Suto, Y.\ 2010, arXiv:1006.4486 

\bibitem[Reese et al.(2010)]{2010ApJ...721..653R} Reese, E.~D., Kawahara, 
H., Kitayama, T., Ota, N., Sasaki, S., \& Suto, Y.\ 2010, \apj, 721, 653 








\bibitem[Renzini et al.(1993)]{Renzini1993} Renzini, A., Ciotti, 
L., D'Ercole, A., \& Pellegrini, S.\ 1993, \apj, 419, 52 

\bibitem[Roediger et al.(2007)]{2007MNRAS.375...15R} Roediger, E., 
Br{\"u}ggen, M., Rebusco, P., B{\"o}hringer, H., 
\& Churazov, E.\ 2007, \mnras, 375, 15 



\bibitem[Rossetti 
\& Molendi(2010)]{2010A&A...510A..83R} Rossetti, M., \& Molendi, S.\ 2010, \aap, 510, A83 



%\bibitem[Sato et al.(2007)]{2007PASJ...59..299S} Sato, K., et al.\ 2007, 
%\pasj, 59, 299 
\bibitem[Sato et al.(2008)]{2008PASJ...60S.333S} Sato, K., Matsushita, K., 
Ishisaki, Y., Yamasaki, N.~Y., Ishida, M., Sasaki, S., 
\& Ohashi, T.\ 2008, \pasj, 60, 333 
\bibitem[Sato et al.(2009)]{2009PASJ...61S.365S} Sato, K., Matsushita, K., 
\& Gastaldello, F.\ 2009, \pasj, 61, 365 
\bibitem[Simionescu et 
al.(2008)]{2008A&A...482...97S} Simionescu, A., Werner, N., Finoguenov, A., B{\"o}hringer, H., \& Br{\"u}ggen, M.\ 2008, \aap, 482, 97 



\bibitem[Simionescu et 
al.(2009)]{2009A&A...493..409S} Simionescu, A., Werner, N., B{\"o}hringer, H., Kaastra, J.~S., Finoguenov, A., Br{\"u}ggen, M., \& Nulsen, P.~E.~J.\ 2009, \aap, 493, 409 




\bibitem[Smith et al. 2001]{apec} Smith, R.K., Brickhouse, N.S., Liedahl,D.A., \& Raymond,
        J.S., 2001, ApJ, 556, 91
\bibitem[\protect\citeauthoryear{Snowden et
		   al.}{2008}]{2008A&A...478..615S} Snowden S.~L.,
			   Mushotzky R.~F., Kuntz K.~D., Davis D.~S.,
			   2008, A\&A, 478, 615 
\bibitem[Tanaka et al.(1994)]{1994PASJ...46L..37T} Tanaka, Y., Inoue, H., 
\& Holt, S.~S.\ 1994, \pasj, 46, L37

\bibitem[Tamura et al.(2004)]{2004A&A...420..135T} Tamura, T., Kaastra, 
J.~S., den Herder, J.~W.~A., Bleeker, J.~A.~M., \& Peterson, J.~R.\ 2004, 
\aap, 420, 135 

\bibitem[Tawa (2008)]{tawa2008} Tawa, N.,2008,  phD thesis, University of Osaka
\bibitem[Vikhlinin et al.(2005)]{2005ApJ...628..655V} Vikhlinin, A., 
Markevitch, M., Murray, S.~S., Jones, C., Forman, W., 
\& Van Speybroeck, L.\ 2005, \apj, 628, 655 

\bibitem[Werner et 
al.(2006)]{2006A&A...449..475W} Werner, N., de Plaa, J., Kaastra, J.~S., Vink, J., Bleeker, J.~A.~M., Tamura, T., Peterson, J.~R., \& Verbunt, F.\ 2006, \aap, 449, 475 



\bibitem[Yoshino et al.(2009)]{2009PASJ...61..805Y} Yoshino, T., et al.\ 
2009, \pasj, 61, 805 
W

\bibitem[Zhang et al.(2009)]{2009ApJ...699.1178Z} Zhang, Y.-Y., Reiprich, 
T.~H., Finoguenov, A., Hudson, D.~S., 
\& Sarazin, C.~L.\ 2009, \apj, 699, 1178 


\end{thebibliography}
\end{document}